\documentclass{article}
\usepackage[T1]{fontenc}
\usepackage[utf8]{inputenc}

\usepackage[english]{babel}
\usepackage{csquotes}

\usepackage{lscape}
\usepackage{amsmath}
\usepackage{mathtools}
\usepackage{amsfonts}
\usepackage{amssymb}
\usepackage{algorithm}
\usepackage{algpseudocode}
\usepackage{enumerate}
\usepackage{physics}
\usepackage{hyperref}
\usepackage{todonotes}
\usepackage{bbm}
\usepackage{longtable}
\usepackage{url}
\usepackage{color}
\usepackage{graphicx}
\usepackage{epstopdf}
\usepackage{multirow}
\usepackage{bm}

\usepackage[backend = biber, style = apa, natbib]{biblatex}
\usepackage{verbatim}
\usepackage{float}
\usepackage[toc,page]{appendix}
\usepackage[onehalfspacing]{setspace}
\usepackage{adjustbox}
\usepackage{rotating}
\usepackage{amsthm}
\usepackage{booktabs,dcolumn,caption}
\usepackage{tikz}

\usepackage{array}

\usepackage{booktabs,caption}
\usepackage{subcaption}
\let\tilde\widetilde

\makeatletter
\newcommand*\rel@kern[1]{\kern#1\dimexpr\macc@kerna}
\newcommand*\widebar[1]{%
  \begingroup
  \def\mathaccent##1##2{%
    \rel@kern{0.8}%
    \overline{\rel@kern{-0.8}\macc@nucleus\rel@kern{0.2}}%
    \rel@kern{-0.2}%
  }%
  \macc@depth\@ne
  \let\math@bgroup\@empty \let\math@egroup\macc@set@skewchar
  \mathsurround\z@ \frozen@everymath{\mathgroup\macc@group\relax}%
  \macc@set@skewchar\relax
  \let\mathaccentV\macc@nested@a
  \macc@nested@a\relax111{#1}%
  \endgroup
}
\makeatother

\makeatletter
\def\FirstLetterUppercase#1{\expandafter\FirstLetterUppercase@i#1 \@nil}
\def\FirstLetterUppercase@i#1#2 #3\@nil{%
  \MakeUppercase{#1}#2
  \ifx\relax#3\relax\def\next@i{}\else\def\next@i{\expandafter\FirstLetterUppercase@i#3\@nil}\fi%
  \next@i}
\makeatother
\renewbibmacro*{title}{\printtext{\printfield{title}}\newunit}

\newcolumntype{L}[1]{>{\raggedright\let\newline\\\arraybackslash\hspace{0pt}}p{#1}}
\newcolumntype{C}[1]{>{\centering\let\newline\\\arraybackslash\hspace{0pt}}p{#1}}
\newcolumntype{R}[1]{>{\raggedleft\let\newline\\\arraybackslash\hspace{0pt}}p{#1}}

\newcommand{\yc}[1]{{{\color{black}  #1}}}
\newcommand{\yx}[1]{{{\color{black}  #1}}}

\newcommand{\clo}[1]{\widebar{\mathcal{#1}}}
\def\bf#1{\mathbf{#1}}

\newcommand{\ttt}{\boldsymbol \theta}

\newcommand{\OO}{\mbox{$\mathbf O$}}

\newcommand{\XX}{\mathbf X}

\newcommand{\XXko}{\tilde{\mathbf{X}}^{\text{obs}}}
\newcommand{\XXu}{\mathbf{X}^\ast}

\newcommand{\SSigma}{\mathbf{\Sigma}}

\newcommand{\argmax}{\operatornamewithlimits{arg\,max}}
\newcommand{\argmin}{\operatornamewithlimits{arg\,min}}

\newtheorem{theorem}{Theorem}

\newtheorem{corollary}{Corollary}

\newtheorem{lemma}{Lemma}

\newtheorem{proposition}{Proposition}
\newtheorem{remark}{Remark}

\title{Rotation to Sparse Loadings using $L^p$ Losses and \\
Related Inference Problems}
\author{Xinyi Liu, Gabriel Wallin, Yunxiao Chen, and Irini Moustaki\\
London School of Economics and Political Science}
\date{} 
\begin{document}

\maketitle
	\begin{abstract}

Researchers have widely used exploratory factor analysis (EFA) to learn the latent structure underlying multivariate data. Rotation and regularised estimation are two classes of methods in EFA that they often use to find interpretable loading matrices. In this paper we propose a new family of oblique rotations based on component-wise $L^p$ loss functions $(0 < p\leq 1)$ that is closely related to an $L^p$ regularised estimator. We develop model selection and post-selection inference procedures based on the proposed rotation method. When the true loading matrix is sparse, the proposed method tends to outperform traditional rotation and regularised estimation methods in terms of statistical accuracy and computational cost. Since the proposed loss functions are nonsmooth, we develop an iteratively reweighted gradient projection algorithm for solving the optimisation problem. We also develop theoretical results that establish the statistical consistency of the estimation, model selection, and post-selection inference. We evaluate the proposed method and compare it with regularised estimation and traditional rotation methods via simulation studies. We further illustrate it using an application to the Big Five personality assessment.

	\end{abstract}
	Keywords: Component loss function, analytic rotation,  regularised estimation, model selection, confidence interval

\section{Introduction}

Researchers have widely used exploratory factor analysis (EFA) to learn the latent structure underlying multivariate data. A major problem in EFA is identifying an interpretable factor structure among infinitely many equivalent solutions that give the same data distribution, where two equivalent solutions  differ by a rotation transformation \citep[see Chapters 10-12,][]{mulaik2009foundations}. Mathematically, we aim to find a sparse solution for which many entries of the loading matrix are exactly or approximately zero so that each factor can be interpreted based on a small number of manifest variables whose loadings on the factor are not close to zero. This idea dates back to the seminal discussion on simple factor structure in \citet{thurstone1947multiple}. 

We can classify methods for obtaining sparse loading structures into two categories -- rotation and regularised estimation methods. A rotation method involves two steps. In the first, we obtain an estimate of the loading matrix. Typically, but not necessarily, a maximum likelihood estimator is used in this step \citep{bartholomew2011latent},  under some arbitrary but mathematically convenient constraints that avoid rotational indeterminacy. \yc{In the second step, we rotate the estimated loading matrix to minimise a certain loss function  where a smaller loss function value tends to imply a more interpretable  solution.} Researchers have proposed different rotation methods that differ by, first, whether the factors are allowed to be correlated, and second, the loss function for measuring sparsity. A rotation method is called an orthogonal rotation when the factors are constrained \yx{to be} uncorrelated and an oblique rotation otherwise. Different loss functions have been proposed for orthogonal and oblique rotations, including  
varimax \citep{kaiser1958varimax}, oblimin \citep{jennrich1966rotation}, 
geomin \citep{yates1987multivariate}, simplimax \citep{kiers1994simplimax}, and component-wise loss \citep{jennrich2004rotation,jennrich2006rotation}, among many others. %We refer the readers to \citet{browne2001overview} for a comprehensive review of rotation methods.  
Among the existing rotation methods, we draw attention to the monotone concave Component Loss Functions \citep[CLFs;][]{jennrich2004rotation,jennrich2006rotation} due to their desired theoretical properties and superior performance in recovering sparse loading matrices. Specifically,  \cite{jennrich2004rotation,jennrich2006rotation}
provided some theoretical guarantees to the CLFs when the true loading matrix has a perfect simple structure and further found that the CLFs are often more accurate in recovering sparse loading matrices than other rotation methods under both orthogonal and oblique settings. 

In recent years, several regularised estimation methods have been proposed for EFA \citep[e.g.,][]{trendafilov2014simple, yamamoto2017graphical, jin2018approximated, geminiani2021single}. 
Slightly different from rotation methods, 
a regularised estimation method simultaneously estimates the model parameters and \yc{produces a sparse solution}. It introduces a least absolute shrinkage and selection operator \citep[LASSO;][]{tibshirani1996regression} type \yx{sparsity}-inducing regularisation term into the loss function for parameter estimation, where the regularisation term  imposes sparsity on the estimated loadings. It typically obtains a sequence of candidate models by varying the weight of the regularisation term in the loss function. The final model is chosen from the candidate models, often using an information criterion. 

In this paper we propose a new family of oblique 
rotations based on component-wise $L^p$ loss functions, for $0 < p\leq 1$. We show the proposed loss functions to be special cases of monotone concave CLFs and that they thus share the same theoretical properties. We note that \citet{jennrich2004rotation,jennrich2006rotation} considered the $L^1$ loss function but not the 
$L^p$ loss functions with  $p<1$. With the proposed rotations, we solve several previously unaddressed problems regarding rotation and regularised estimation methods. First, we establish the statistical consistency of the rotated solution. More specifically, we provide conditions under which the rotated solution converges to the true sparse loading matrix as the sample size goes to infinity. These conditions also provide insights into the choice of $p$.
Seemingly straightforward, this consistency result requires some refined analysis and, to our best knowledge, such results have not been established for other rotation methods. In particular, the theoretical results for the CLFs in \citet{jennrich2004rotation,jennrich2006rotation} were established concerning the population loading matrix rather than its estimate. Second, we address the difficulty of establishing whether regularised estimation methods outperform rotation methods or vice versa. To gain some insights into this question, we theoretically show  that the proposed rotation method can be viewed as the limiting case of a
regularised estimator when the weight of the regularisation term converges to zero.  In addition, 
to compare the two methods in terms of model selection, we develop a hard-thresholding procedure that conducts 
model selection based on a rotated solution. Through computational complexity analysis and simulation studies, we find that the proposed method  achieves similar statistical accuracy
as regularised estimation given a reasonable sample size and is computationally faster. 
Third, monotone concave CLFs, including the proposed $L^p$ loss functions, are not smooth everywhere. Consequently, the traditional gradient projection algorithms  are no longer applicable. 
\citet{jennrich2004rotation,jennrich2006rotation}
bypassed the computational issue by replacing a CLF with a  smooth approximation and pointed out potential issues with this treatment. 
We propose an Iteratively Reweighted Gradient Projection (IRGP) algorithm that may better solve this nonsmooth optimisation problem.  Finally, uncertainty quantification for the rotated solution 
affects the interpretation of the factors and, thus, is vital in EFA. However, the delta method, which is used to obtain confidence intervals for rotation methods with a smooth objective function \citep{jennrich1973standard}, is not applicable due to the nonsmoothness of the current loss functions. That is, the delta method requires the loss function to be smooth at the true loading matrix, which is not satisfied for monotone concave CLFs. We tackle this problem by developing a post-selection inference procedure that gives asymptotically valid confidence intervals for loadings in a rotated solution.  We evaluate the proposed method and compare it with regularised estimation and traditional rotation methods via simulation studies. We further illustrate it using an application to the Big Five personality assessment.

The rest of the paper is structured as follows. In Section~\ref{sec:meth} we propose $L^p$ criteria for oblique rotation, and draw a connection with regularised estimation. In Section~\ref{sec:inf} we discuss statistical 
inferences based on the proposed rotation method and establish their asymptotic properties, and in Section~\ref{sec:comp} we develop an iteratively reweighted gradient projection algorithm for solving the optimisation problem associated with the proposed rotation criteria. We evaluate the proposed method via simulation studies in Section~\ref{sec:sim} and an application to the  Big Five personality assessment in Section~\ref{sec:real}. We conclude this paper with discussions on the limitations of the proposed method and future directions in Section \ref{sec:conc}. Proof of the theoretical results, additional 
 simulation results, and further details of the real application 
are given in the supplementary material. 

%\yc{Add the organisation of the paper once the paper is done.}

\iffalse
Section 2 introduces the $L^p$ criteria for oblique rotation. It furthermore presents a theoretical result connecting the obtained solution under the $L^p$ rotational family with the regularized estimation solution. Section 3 outlines the hard-thresholding procedure used for model selection and presents the statistical consistency of the proposed estimator. In Section 4, procedures for drawing statistical inference on the sparse true loading matrix are developed. This is in Section 5 followed by a description of the computational methods developed for the optimization problem. Section 6 presents a simulation study and Section 7 introduces a case study based on the Big 5 data set. The paper concludes with a discussion. All proofs are found in the appendix.
\fi

\section{\texorpdfstring{$L^p$}{} Rotation Criteria}\label{sec:meth}
\subsection{Problem Setup} 
%\textcolor{red}{
%First, set up the model. 
%Second, describe the two-step procedure of rotation methods. Step 1, obtain an orthogonal solution that maximises the likelihood function. Step 2, apply rotation to the loading matrix of the orthogonal solution. You can describe both the orthogonal and oblique rotations (give their definitions) here, but say that we will focus on the 	oblique rotation here, 	as it is more general. }

We consider an exploratory linear factor model with $J$ indicators and $K$ factors given by
\begin{equation}\label{eq:1}
    \mathbf{X}|\boldsymbol{\xi} \sim \mathcal{N}( \boldsymbol{\Lambda}\boldsymbol{\xi} , \mathbf{\Omega} ),
\end{equation}
where $\mathbf{X}$ is a $J$-dimensional vector of manifest variables, $\boldsymbol{\Lambda} = (\lambda_{jk})_{J\times K}$ is the loading matrix,  $\boldsymbol{\xi}$ is a $K$-dimensional vector of common factors, and $\boldsymbol\Omega = (\omega_{ij})_{J\times J}$ denotes the residual covariance matrix. It is assumed that the common factors are normally distributed with variances fixed to 1, i.e, 
$
\boldsymbol{\xi} \sim \mathcal{N}(\boldsymbol{0},\boldsymbol{\Phi}),
$
where $\boldsymbol{\Phi}\in \mathbb{R}^{K \times K}$ has diagonal entries $\phi_{kk}$, $k = 1, \ldots, K$, equal to 1 and is symmetric positive definite, denoted by $\boldsymbol{\Phi} \succ 0$. The manifest variables are assumed to be conditionally independent given $\boldsymbol{\xi}$, i.e., the off-diagonal entries of $\mathbf{\Omega}$ are set to 0. To simplify the notation, we use
$\boldsymbol{\theta} = (\boldsymbol{\Lambda}, \boldsymbol{\Phi}, \boldsymbol{\Omega})$ to denote all of the unknown parameters. The model in \eqref{eq:1} implies the marginal distribution of $\XX$
\begin{equation}\label{eq:2}
\mathbf{X} \sim \mathcal{N}(\boldsymbol{0},\boldsymbol{\Sigma(\theta)}),
\end{equation}
where 
$
\boldsymbol{\Sigma(\theta)} = \boldsymbol{\Lambda} \boldsymbol{\Phi} \boldsymbol{\Lambda}' + \boldsymbol{ \Omega}.
$
Without further constraints, the parameters in \eqref{eq:2} are not identifiable due to rotational indeterminacy. That is, two sets of parameters $\ttt$ and 
$\tilde\ttt = (\tilde{\boldsymbol{\Lambda}}, \tilde{\boldsymbol{\Phi}}, \tilde{\boldsymbol{\Omega}})$ give the same distribution for $\XX$ if 
$\boldsymbol{\Lambda} \boldsymbol{\Phi} \boldsymbol{\Lambda}' = \tilde{\boldsymbol{\Lambda}} \tilde{\boldsymbol{\Phi}} \tilde{\boldsymbol{\Lambda}}'$
and 
$\boldsymbol{\Omega} = \tilde{\boldsymbol{\Omega}}$. Note that the normality assumptions above are not essential. We adopt them for ease of writing, and the development in the current paper does not rely on these normality assumptions. Throughout this paper, we assume that the number of factors $K$ is known. 

%Due to the rotational indeterminacy, one can replace $\boldsymbol{\Phi}$ and $\boldsymbol{\Lambda}$ by 

An oblique rotation method is a two-step procedure. In the first step, one obtains an estimate of the model parameters, under the constraints that $\boldsymbol{\Phi} = \mathbf I$ and other arbitrary but mathematically convenient constraints that fix the rotational indeterminacy. Note that due to the rotational indeterminacy, we can always constrain $\boldsymbol{\Phi} = \mathbf I$ and absorb the dependence between the factors into the loading matrix $\boldsymbol \Lambda$. We can obtain the estimate using any reasonable estimator for factor analysis, such as the least-square \citep{joreskog1972factor}, weighted-least-square \citep{browne1984asymptotically}, and maximum likelihood estimators \citep{joreskog1967some}. We denote this estimator by $\hat{\boldsymbol{\theta}} = (\hat{\mathbf{A}}, \mathbf{I},\hat{\boldsymbol{\Omega}})$.  In the second step, we find an oblique rotation matrix $\hat{\mathbf T}$, such that the rotated loading matrix $\hat{\mathbf{\Lambda}}= \hat{\mathbf{A}} \mathbf{\hat{T}}^{{\prime}-1}$ minimises a certain loss function $Q$ that measures the sparsity level of a loading matrix. We will propose the functional form of $Q$ in the sequel. Here, an oblique rotation matrix $\mathbf T$ satisfies that $\mathbf T$ is invertible and  $(\mathbf{T'} \mathbf{T})_{kk} = 1, \, k=1, \ldots, K$. Consequently, any rotated solution $(\hat{\mathbf{A}} \mathbf{ {T}}^{{\prime}-1}, \mathbf{T'} \mathbf{T}, \hat{\boldsymbol{\Omega}})$ is still in the parameter space and gives the same distribution for $\XX$. More precisely, we let 
\begin{align}\label{eq:13}
\mathcal{M}=&\{ \mathbf{T} \in \mathbb{R}^{K \times K} : rank(\mathbf{T}) = K, \, (\mathbf{T'} \mathbf{T})_{kk} = 1, \, k=1, \ldots, K \}
\end{align} 
be the space for oblique rotation matrices, where $rank(\cdot)$ gives the rank of a matrix. Then the oblique rotation problem involves solving the optimisation 
\begin{equation}\label{eq:rotation}
\mathbf{\hat{T}} \in \argmin_{ \mathbf{T} \in \mathcal{M}} Q(\hat{\mathbf{A}} \mathbf{T'}^{-1}),
\end{equation}
and the rotated solution is given by $(\hat{\mathbf{A}} \hat{\mathbf{{T}}}^{{\prime}-1}, \hat{\mathbf{T}}' \hat{\mathbf{T}}, \hat{\boldsymbol{\Omega}})$. Equivalently, the rotated loading matrix $\hat{\boldsymbol\Lambda}$ satisfies 

\begin{equation}\label{eq:rotation2}
  (\hat{\boldsymbol\Lambda}, \hat{\boldsymbol\Phi}) \in \argmin_{\boldsymbol\Lambda, \boldsymbol\Phi} Q(\boldsymbol\Lambda), \mbox{~such that~} \boldsymbol\Lambda \boldsymbol\Phi\boldsymbol\Lambda' = \hat{\mathbf A}\hat{\mathbf A}', \boldsymbol{\Phi} \succ 0, \mbox{~and~}\phi_{kk}=1, k = 1, \ldots, K.
\end{equation}

As explained in Remark~\ref{rmk:indet}, the minimiser of \eqref{eq:rotation}, or equivalently that of \eqref{eq:rotation2}, is not unique. 

\begin{remark}\label{rmk:indet}
Let $\mathcal{D}_1$ be the set of all $K\times K$ permutation matrices and $\mathcal{D}_2$ be the set of all $K\times K$ sign flip matrices. 
For any  $\mathbf D_1 \in \mathcal{D}_1$, $\mathbf D_2\in \mathcal D_2$, and $K\times K$ matrix $\mathbf T$, 
$\mathbf T \mathbf D_1$ is a matrix whose columns are a permutation of those of $\mathbf T$ and, 
$\mathbf T\mathbf D_2$ is a matrix whose $k$th column is either the same as the $k$th column of $\mathbf T$ or  the $k$th column of $\mathbf T$ multiplied by $-1$.  Let $\hat{\mathbf T}$ be one solution to the optimisation problem   \eqref{eq:rotation}. It is easy to check that $\hat{\mathbf T} \mathbf  D_1 \mathbf  D_2$ also minimises the objective function \eqref{eq:rotation}, for any $\mathbf  D_1\in \mathcal{D}_1$ and $\mathbf  D_2\in \mathcal{D}_2$. The resulting loading matrix is equivalent to 
$\hat{\boldsymbol\Lambda}$ up to a column permutation and column sign flips. 

\end{remark}

We conclude the problem setup with two remarks. 

\begin{remark}
The rotation problem not only applies to the linear factor model, but also other settings, such as item factor analysis \citep{reckase2009multidimensional, chen2019joint, chen2021item} and machine learning models such as the stochastic blockmodel and latent Dirichlet allocation \citep[see][]{rohe2020vintage}. \yc{These models are all latent variable models involving  manifest variables $\XX$, latent variables $\boldsymbol\xi$,  a parameter matrix $\boldsymbol\Lambda$, and possible other model parameters. The parameter matrix $\boldsymbol\Lambda$ 
connects $\XX$ and $\boldsymbol\xi$, 
playing a similar role to the loading matrix in the linear factor model. 
We can view these models as extensions of the linear factor model to more general variable types (e.g., binary or categorical) with more flexible assumptions on the distribution of $(\XX, \boldsymbol\xi)$. We can apply the rotation method to learn an interpretable $\boldsymbol\Lambda$ in these models.}

%\yx{The multidimensional item response theory model, popular in educational and psychological measurement, is an example of the former and could be considered a special case of the factor model studied in this paper but with categorical outcome. In \citet[][]{rohe2020vintage}, connections between the factor analysis model and popular machine learning models such as the Stochastic blockmodel and Latent Dirichlet Allocation are established.}
\end{remark}

\begin{remark}

Although in the current paper we focus on oblique rotations, we note that the proposed criteria can be easily extended to orthogonal rotation, as the latter can be viewed as a special case of the former when $\boldsymbol\Phi$ is fixed to be an identify matrix. That is, given a loss function $Q$, orthogonal rotation solves the problem 

$$\min_{\boldsymbol\Lambda} Q(\boldsymbol\Lambda), \mbox{~such that~} \boldsymbol\Lambda \boldsymbol\Lambda' = \hat{\mathbf A}\hat{\mathbf A}'.$$

%\yc{Give the orthogonal rotation setting and show that it is a special case of the oblique rotation. Explain that the development in the current paper can be easily adapted to the orthogonal rotation problem.}
\end{remark}

\iffalse
which refer to the problem that $ (\boldsymbol{\Lambda}(\mathbf{T'})^{-1},\mathbf{T}'\mathbf{T} ,\mathbf{\Omega} ) $ and $ (\boldsymbol{\Lambda},\mathbf{I},\mathbf{\Omega})$ result in the same distribution for $\mathbf{X}$ as long as $\mathbf{T}'\mathbf{T} $ is a vaild candidate for $\boldsymbol{\Phi}$, i.e., when the diagonal entries of $\mathbf{T}'\mathbf{T}$ all take value 1. On the other hand, this gives us the opportunity to find a simpler and more interpretable structure, assuming that each manifest variable is accounted by only a few common factors. That is to say, the true loading matrix is sparse.
\fi
%Although debated what qualifies as a simple structure, a %perfect simple structure equals at most one non-zero element %per row in the loading matrix. 
%Typically, either orthogonal (i.e, %$\boldsymbol{\Phi}=\mathbf{I}$ ) or oblique rotation (i.e, %$\mathbf{\Phi} \succ 0,\boldsymbol{\Phi}_{ii}=1$, for $i %=1,2,..,k$) is considered based on whether or not the final %solution permits correlation between the common factors. 

%We denote the obtained estimate by $$

\subsection{Proposed Rotation Criteria}\label{subsec:propose}

\citet{jennrich2004rotation,jennrich2006rotation} proposed a family of monotone concave CLFs for the choice of $Q$ in \eqref{eq:rotation}, taking the form 

%As for the choice of rotation criterion $Q(\boldsymbol{\Lambda})$, we study the class of concave component loss functions (CLCs) suggested in \citet{jennrich2002simple}, 
%
\begin{equation}\label{eq:CLC_jennrich}
Q(\mathbf{\Lambda}) =\sum_{j=1}^J\sum_{k=1}^K h(|\lambda_{jk}|),    
\end{equation}
where $\mathbf{\Lambda} = (\lambda_{jk})_{J\times K}$ and $h$ is a concave and monotone increasing function
that maps from $[0,\infty)$ to $[0,\infty)$. 
This family of loss functions is appealing for several reasons. First, a CLF takes a simple form that does not involve products of loadings and their higher-order polynomial terms. Second, the monotone concave CLFs have desirable properties. In particular, \citet{jennrich2006rotation} proved that a monotone concave CLF is minimised by loadings with a perfect simple structure when such \yx{a} loading structure \yx{exists}.  Third, simulation studies in 
\citet{jennrich2004rotation,jennrich2006rotation} showed that these loss functions tend to outperform traditional rotation methods (e.g., promax, simplimax, quartimin, and geomin) when the true loading matrix is sparse. 

Two examples of $h$ are given in  \citet{jennrich2004rotation,jennrich2006rotation}, 
including the linear CLF 
where $h(|\lambda|) = |\lambda|$ 
and the basic CLF 
where $h(|\lambda|) = 1 - \exp(-|\lambda|)$. However, there does not exist a full spectrum of monotone concave CLFs for dealing with true loading matrices with different sparsity levels. 
To fill this gap, we propose a general family of 
monotone concave CLFs that we name the $L^p$ CLFs. More specifically, for each value of $p \in (0,1]$, the loss function takes the form

\begin{equation}\label{eq:CLC_Lp}
Q_p(\mathbf{\Lambda}) =\sum_{j=1}^J\sum_{k=1}^K |\lambda_{jk}|^p.
\end{equation}
Proposition~\ref{prop:concave} below shows that this choice of $h$ yields a monotone concave CLF. 

\begin{proposition}\label{prop:concave}
The absolute value function $h(x) = |x|^p$, $p \in (0, 1]$ is monotonically increasing and concave on the interval $[0, \infty)$.

%A proposition stating that $h(x) = |x|^p$ is monotone increasing and concave on $[0,\infty)$.}
\end{proposition}

Under very mild regularity conditions, any $L^p$ CLF is uniquely minimised by a loading matrix of perfect simple structure, when such a loading matrix exists, where we say the minimiser is unique when all the minimisers of the loss function are equivalent up to column permutation and sign flip transformations (see Remark~\ref{rmk:indet} for these transformations). 
%the loss function is minimised by a specific loading matrix or any 
We summarise this result in Proposition~\ref{prop:simple} below. This result improves Theorem~1 of \citet{jennrich2006rotation}, as the uniqueness of the perfect simple structure is not established in \citet{jennrich2006rotation} for the $L^1$-criterion. 

%According to Proposition~\ref{prop:concave} and the theoretical results from \citet{jennrich2006rotation}, any $L^p$ CLF is minimised by loadings with perfect simple structure when such loadings exist. We summarise this result in Proposition~\ref{prop:simple} below. The proof of these propositions is given as supplementary materials. 

\begin{proposition}\label{prop:simple}
Suppose that the true loading matrix $\boldsymbol\Lambda^*$ has perfect simple structure, in the sense that each row has at most one non-zero entry. Further suppose that $\boldsymbol\Lambda^*$ is of full column rank, i.e.,  $rank(\boldsymbol\Lambda^*)=K$. Then, for any oblique rotation matrix $\mathbf T\in \mathcal{M}$, 
$$Q_p(\boldsymbol\Lambda^* \mathbf{T'}^{-1}) \geq Q_p(\boldsymbol\Lambda^*),$$
where the two sides are equal if, and only if, 
$\mathbf{T'}^{-1} = \mathbf D_1\mathbf D_2$  for  $\mathbf D_1 \in \mathcal{D}_1$ and $\mathbf D_2\in \mathcal D_2$; see Remark~\ref{rmk:indet} for the definitions of $\mathcal{D}_1$ and $\mathcal D_2$.

%If there is an oblique rotation $\hat{\bs\Lambda}$ of $\bf A$ that has perfect simple structure, we have (1) If $p \leq 1$, then $\hat{\bs\Lambda} \in \{ \mathbf{A} \mathbf{T'}^{-1} : \mathbf{T} \in \argmin_{\mathbf{T} \in \mathcal{M}} Q_p(\mathbf{A} \mathbf{T'}^{-1}) \} $, i.e, $\hat{\bs\Lambda}$ minimises $Q_p(\mathbf{\Lambda})$ over all oblique
%rotations of $\bf A$. (2) If $p < 1$, any elements of $\{ \mathbf{A} \mathbf{T'}^{-1} : \mathbf{T} \in \argmin_{\mathbf{T} \in \mathcal{M}} Q_p(\mathbf{A} \mathbf{T'}^{-1}) \} $ must have perfect simple structure.
%\yx{(The ``moreover" part of the statement is not clear. Also, I felt that you need to write this statement in a more mathematical way.)}
\end{proposition}

\begin{figure}[t]  
\centering
     \begin{subfigure}{.45\textwidth} 
         \centering
         \includegraphics[width=.9\linewidth]{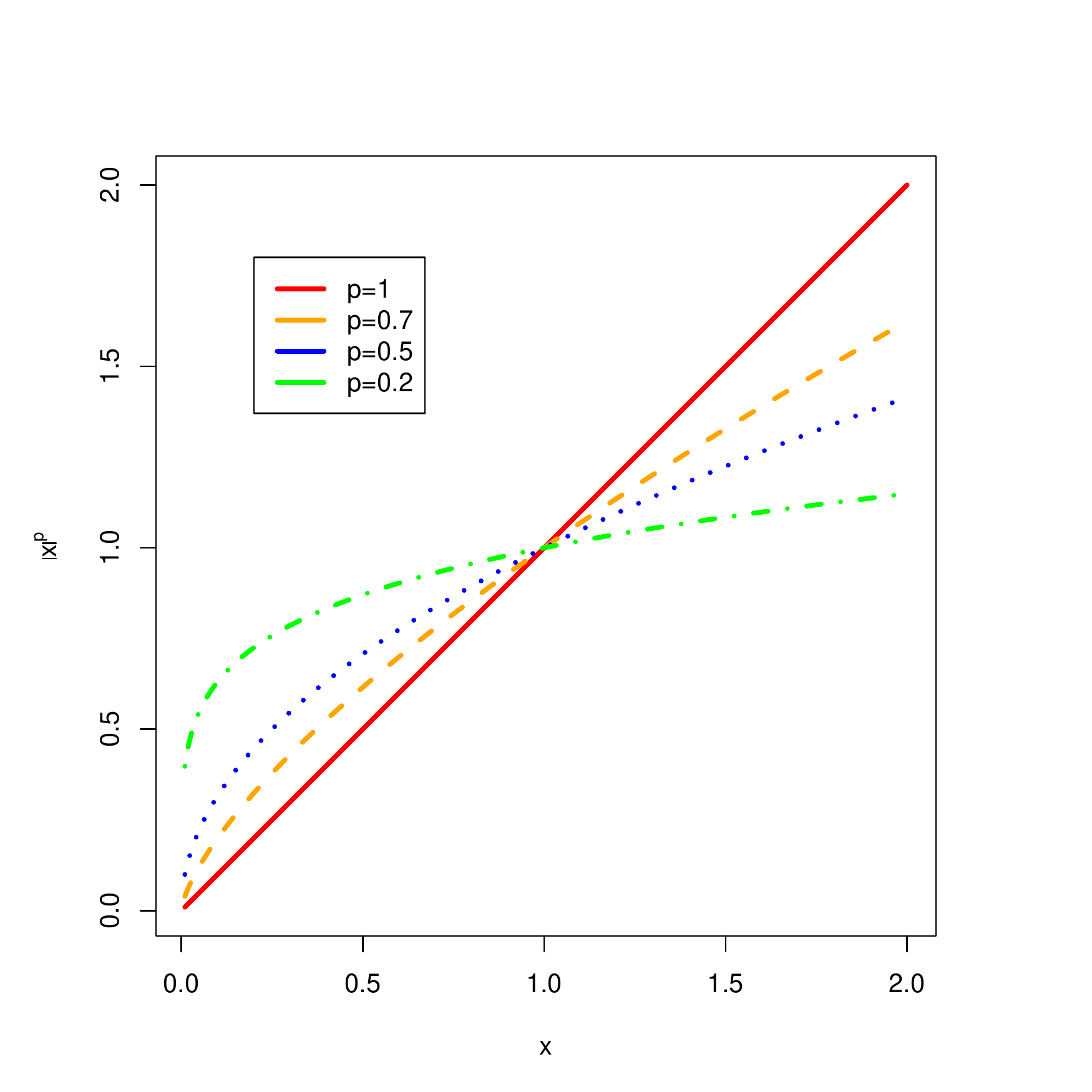}
         \caption{}
         %\label{fig:p=0.5}
     \end{subfigure}
     \hfill
     \begin{subfigure}{.45\textwidth}
         \centering
         \includegraphics[width=.9\linewidth]{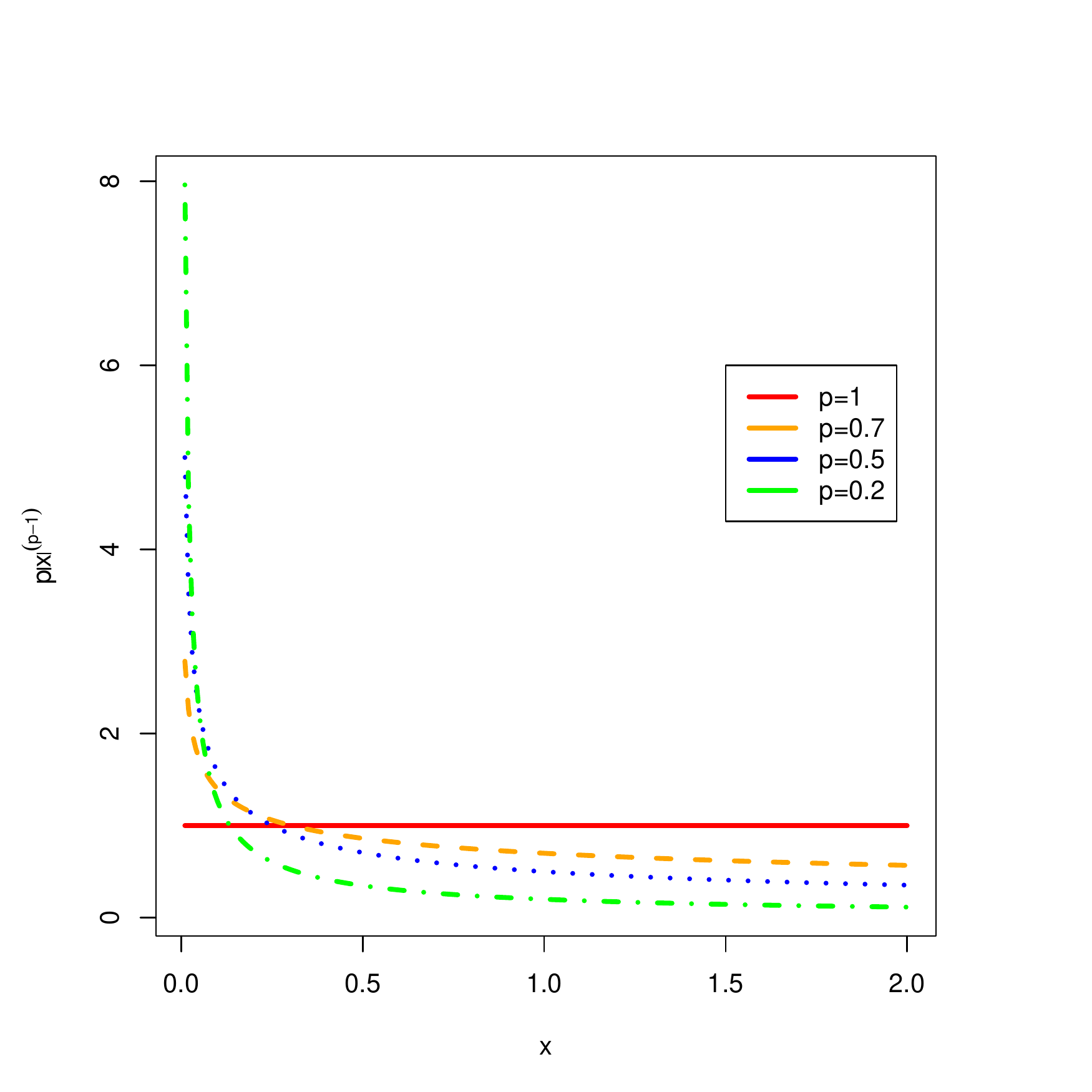}
         \caption{}
         %\label{fig:p=1}
     \end{subfigure}
\caption{Panel (a): Plots of $|x|^p$, for different choices of $p$. Panel (b): Plots of the derivative of  $|x|^p$, for different choices of $p$. }
\label{fig:lp}
\end{figure}

\begin{figure}[t]  
\centering
     \begin{subfigure}{.45\textwidth} 
         \centering         \includegraphics[width=.9\linewidth]{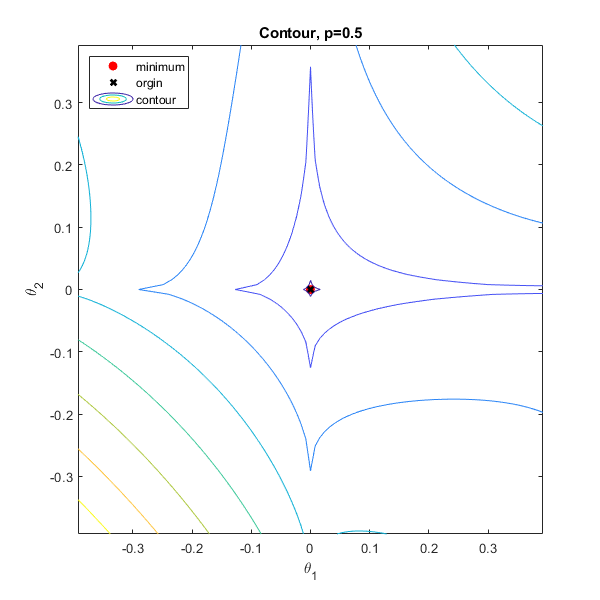}
         \caption{}
         %\label{fig:p=0.5}
     \end{subfigure}
     \hfill
     \begin{subfigure}{.45\textwidth}
         \centering
         \includegraphics[width=.9\linewidth]{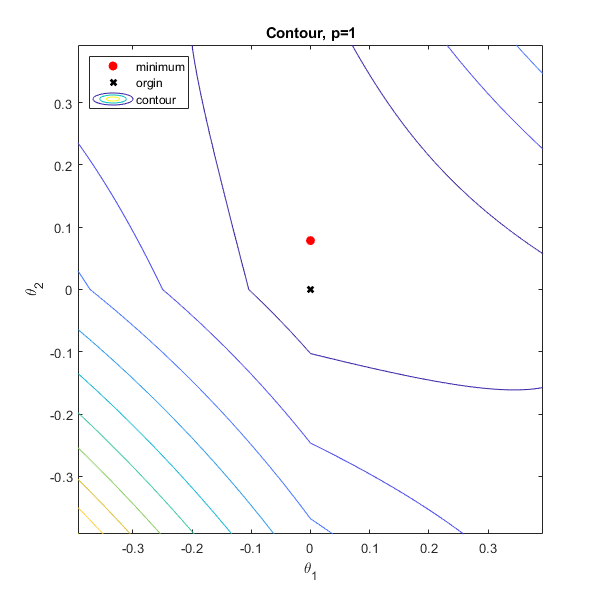}
         \caption{}
         %\label{fig:p=1}
     \end{subfigure}
        \caption{Plots of contours of $|\boldsymbol{\Lambda}^{*}\mathbf{T}^{-1'}|^p$, where $\mathbf{T}=[ \cos(\theta_1),\sin(\theta_2);\sin(\theta_1),\cos(\theta_2)]$. Panel (a): $p=0.5$. Panel (b): $p=1$.}
        \label{fig:contour}
\end{figure}

Why do we need the loss functions with $p <1$, given that the choice of $p=1$ is already available in \citet{jennrich2004rotation,jennrich2006rotation}? This is because different $L^p$ CLFs may behave differently when the true loading matrix  does not have a perfect simple structure but still contains many zero loadings. Such a loading structure is more likely to be recovered by an $L^p$ CLF when $p<1$ than by the $L^1$ CLF. 
In what follows, we elaborate on this point. Let 
$\mathbf{\Lambda}^*$ be the true sparse loading matrix and $\boldsymbol{\Phi}^*$ be the corresponding covariance matrix for the factors. For the true loading matrix $\mathbf{\Lambda}^*$ to be recovered by an $L^p$ CLF, a minimum requirement is that 
\begin{equation}\label{eq:minimum}
Q_p(\mathbf{\Lambda}^*) = \min_{\mathbf{\Lambda}} Q_p(\mathbf{\Lambda}), \mbox{~such that there exists~}  \boldsymbol{\Phi} \succ 0, \phi_{kk}=1, k = 1, ..., K,  \boldsymbol{\Lambda}^* \boldsymbol{\Phi}^* \boldsymbol{\Lambda}^{*'} = \boldsymbol{\Lambda} \boldsymbol{\Phi} \boldsymbol{\Lambda}'.
\end{equation}
In other words, $\boldsymbol{\Lambda}^*$ needs to be a stationary point of $Q_p$.  
In Figure~\ref{fig:lp} we give the plots for $|x|^p$ with different choices of $p$ and their derivatives when $x > 0$. 
We note that when $p<1$ the derivative of $|x|^p$ converges to infinity as $x$ approaches zero. The smaller the value of $p$, the faster the convergence speed is. 
On the other hand, when $p=1$, the derivative of $|x|$ takes the value one for any $x > 0$. Therefore,  when $\boldsymbol{\Lambda}^*$ is sparse but does not have a perfect simple structure, it is more likely to be a stationary point of $Q_p$ for $p<1$ than $Q_1$. We illustrate this point by a numerical example, where  
$$
(\boldsymbol{\Lambda}^{*})' = \left(\begin{array}{ccccccc}

    1.20&    0 &   0.15&    0&    0.25&    1.05&    0.18 \\
     0&    0.27&    0&    1.04&    0.15&    1.29&    0.11
\end{array} \right) 
$$
and 
$\boldsymbol{\Phi}^*$ is set to be an identity matrix.
\yc{Note that a $2\times 2$ oblique rotation matrix can be reparameterised by 
$$\mathbf{T}(\theta_1, \theta_2) = \left(\begin{array}{cc}
   \cos(\theta_1)  & \sin(\theta_2) \\
    \sin(\theta_1) & \cos(\theta_2)
\end{array}\right)$$
for $\theta_1, \theta_2 \in [0,2\pi)$.
In Figure~\ref{fig:contour} we show the contour plots of $Q_p(\mathbf{\Lambda}^* \mathbf{T}(\theta_1, \theta_2))$, with $p = 0.5$ and $1$, respectively.}
%It is not difficult to show that when $K=2$, \yx{ the oblique matrix in $\mathcal{M}$ as defined in (\ref{eq:13}) has two degrees of freedoms,  which are parameterised with $(\theta_1, \theta_2)$ by $\mathbf{T}=[ \cos(\theta_1),;,$  in each plot. }
The point $(0,0)$, which is indicated by a black cross, 
corresponds to $\mathbf{\Lambda} =\mathbf{\Lambda}^*$, and the point indicated by a red point  corresponds to the $\mathbf{\Lambda}$ matrix such that $Q_p(\mathbf{\Lambda})$ is minimised.  As we can see, when $p=0.5$, the loss function is minimised by $\mathbf{\Lambda}^*$. On the other hand, when $p=1$, the minimiser of the loss function is not $\mathbf{\Lambda}^*$ and the resulting solution 
does not contain as many zeros as $\mathbf{\Lambda}^*$.   

We emphasise that due to the singularity of the $L^p$ function near zero when $p<1$, the optimisation for $Q_p$ tends to be more challenging with a smaller value of $p$. This is also reflected by the contour plots in Figure~\ref{fig:contour}, where we see  
$Q_{0.5}$ is very non-convex, even around the minimiser. On the other hand, $Q_{1}$ seems locally convex near the minimiser.  Therefore, although the $L^p$-rotation with $p<1$ may be better at recovering sparse loading matrices, its computation is more challenging than the $L^1$-rotation. \yc{Thus, the choice of $p$ involves a trade-off between statistical accuracy and computational cost. We have noticed that despite the above counter example, the $L^1$ criterion tends to give similar results as other $L^p$ criteria ($p<1$) in most simulation and real-data settings that we have encountered. Considering its computational advantage, we recommend users to always start with the $L^1$ criterion. Some smaller $p$ values (e.g., $p=0.5$) may be tried in order to validate the 
$L^1$-rotation result. More guidance on the choice of $p$ can be found in Section \ref{sec:conc}. We discuss the computation of the proposed rotation criteria in Section~\ref{sec:comp}.}

\yc{    
%When solving the optimisation problems, we start with $p=1$ and  use the solution for a larger $p$ value as a warm start when solving the optimisation for a smaller $p$. 

Finally, we remark that when the true loading matrix is sparse but does not have a perfect simple structure, rotation criteria with a smooth objective function (e.g., quartimin and geomin) typically cannot exactly  recover the true sparse loading matrix, even when the true loading matrix can be estimated without error. This is due to the fact that a smooth objective function does not discriminate well between zero parameters and close-to-zero parameters. Thus, such rotation criteria do not favour exactly sparse solutions (i.e., with many zero loadings) and only tend to yield approximately sparse solutions (i.e., many small but not exactly zero loadings). Numerical examples illustrating this point are given in \cite{jennrich2004rotation,jennrich2006rotation}, and a new numerical example and associated simulation results are in Appendix H of the supplementary material.}

%the objective function $Q_p(\boldsymbol\Lambda)$ tends to be more difficult 

\begin{comment}
\begin{figure}[t]
\centering
\includegraphics[width=0.4\textwidth]{L_p.png}
\end{figure}
\end{comment}

%As will be discussed in the sequel, the linear CLF is closely related to a LASSO-regularised estimator for EFA \citep[e.g.,][]{geminiani2021single}. 

 \subsection{Connection with regularised estimation }

The proposed rotation criteria have a close connection with regularised estimators for EFA. In what follows, we establish this connection. Recall that the proposed procedure relies on an initial estimator of the loading matrix for which $\boldsymbol\Phi$ is constrained to be an identity matrix. We further require it to be an $M$-estimator \citep[Chapter 5,][]{van2000asymptotic}, obtained by minimising a certain loss function, denoted by $L(\boldsymbol{\Sigma(\theta)})$. Note that all the popular EFA estimators are $M$-estimators. For instance, when the maximum likelihood estimator is used, then  {the loss function to be minimised is}
$$L(\boldsymbol{\Sigma(\theta)}) = \log \text{det} (\boldsymbol{\Sigma(\theta)}) +  \text{tr}(\boldsymbol{\Sigma(\theta)}^{-1} \boldsymbol{S}),$$
where $
\boldsymbol{S} = (\sum_{i=1}^N \mathbf{x}_i \mathbf{x}_i^\top)/N
$ is the sample covariance matrix. 

Now we introduce an $L^p$ regularised estimator based on the loss function  $L(\boldsymbol{\Sigma(\theta)})$ in the form 
\begin{equation}\label{eq:regul}
\hat \ttt_{\gamma,p} \in \argmin_{\ttt} L(\boldsymbol{\Sigma(\theta)}) + \gamma  \sum_{j=1}^J\sum_{k=1}^K |\lambda_{jk}|^p,
\end{equation}
where $\gamma > 0$ is a tuning parameter and the covariance matrix $\boldsymbol \Phi$ is estimated rather than constrained to be an identity matrix. We note that the minimiser of \eqref{eq:regul} is also not unique due to column permutation and sign flips similar to the non-uniqueness of optimisation \eqref{eq:rotation}. We denote the set of minimisers as  
$$\hat{\mathcal{C}}_{\gamma,p} = \argmin_{\ttt} L(\boldsymbol{\Sigma(\theta)}) + \gamma  \sum_{j=1}^J\sum_{k=1}^K |\lambda_{jk}|^p.$$
Note that the regularisation term takes the same form as the $L^p$ CLF. It is used to impose sparsity on the estimate of the loading matrix. When $p=1$, it becomes a LASSO-regularised estimator that has been considered in, for example, \citet{choi2010penalized}, \citet{hirose2014estimation, hirose2015sparse}, \citet{jin2018approximated}, and \citet{geminiani2021single}.
The regularised estimator \eqref{eq:regul} is similar in spirit to $L^p$-regularised regression \citep{mazumder2011sparsenet,lai2011unconstrained,zheng2017does}, where the $L^p$ regularisation with $p<1$ has been shown to  better recover sparse signals under high-dimensional linear regression settings while computationally more challenging \citep{zheng2017does}. 

%When $0< p < 1$, the regularisation term is 
%known as a bridge regularisation under the regression setting \citep{}. 
%The two terms in \eqref{eq:regul} 
%Let $\hat{\boldsymbol \Lambda}_p$ be the rotated loading matrix given by the $L^p$ CLF, and 

As summarised in Proposition~\ref{prop:regul} below, we can view the proposed $L^p$ rotation solution as a limiting case of the $L^p$-regularised estimator when the tuning parameter $\gamma$ converges to zero.  %The proof of the proposition can be found in the supplementary materials. 

\begin{proposition}\label{prop:regul} 

Consider a fixed $p \in (0,1]$ and a fixed dataset. Suppose that for any sufficiently small $\gamma > 0$, $\hat{\mathcal{C}}_{\gamma,p}$ only contains $n = 2^KK!$ elements that are equivalent up to column permutation and sign flips of the loading matrix, where $K!$ denotes $K$ factorial that counts the number of all possible permutations and $2^K$ gives the total number of sign flip transformations. Furthermore, assume that for any sufficiently small $\gamma > 0$, one can label the elements of $\hat{\mathcal{C}}_{\gamma,p}$, denoted by $\hat{\ttt}_{\gamma, p}^{(i)}$,  $i=1, ..., n$, such that there exists a sufficiently small constant $\delta > 0$, 
$\hat{\ttt}_{\gamma, p}^{(i)}$ is a continuous and bounded function of $\gamma$ in $(0, \delta)$, for each $i$. Then the limit 
$$\hat{\ttt}_{0, p}^{(i)} = (\hat{\boldsymbol\Lambda}_{0,p}^{(i)}, \hat{\boldsymbol\Phi}_{0,p}^{(i)}, \hat{\boldsymbol\Omega}_{0,p}^{(i)}) = \lim_{\gamma \rightarrow 0} \hat{\ttt}_{\gamma, p}^{(i)} $$
exists, and $\hat{\ttt}_{0, p}^{(i)}$ satisfies that 
$(\hat{\boldsymbol\Lambda}_{0,p}^{(i)}, \hat{\boldsymbol\Phi}_{0,p}^{(i)})$ solves the optimisation problem \eqref{eq:rotation2} and  $\hat{\boldsymbol\Omega}_{0,p}^{(i)} = \hat{\boldsymbol\Omega}$, where $\hat{\boldsymbol{\theta}} = (\hat{\mathbf{A}}, \mathbf{I},\hat{\boldsymbol{\Omega}})$ minimises the loss function $L(\boldsymbol{\Sigma(\theta)})$.

\end{proposition}

We now discuss the implications of this connection. First, if we have a numerical solver for the regularised estimator \eqref{eq:regul}, then we can obtain an approximate solution to the $L^p$-rotation problem \eqref{eq:rotation2} by using a sufficiently small tuning parameter $\gamma$.  Second, thanks to this connection, the choice between regularised estimation and rotation becomes the choice of the tuning parameter in regularised estimation. Note that the tuning parameter $\gamma$ corresponds to a bias-variance trade-off in estimating the model parameters $\ttt$. As $\gamma$ increases, the bias of the regularised estimator also increases and the variance decreases. In applications where the sample size is large relative to the number of model parameters, the optimal choice of the tuning parameter is often close to zero.
In that case, it is a good idea to use the rotation method, as the regularised estimator under the optimal tuning parameter may not be substantially more accurate than the rotation solution and searching for the optimal tuning parameter can be computationally costly. \yc{We further discuss this point in a simulation study in Section~\ref{sec:sim}.}
We will discuss the computation of these methods in Section~\ref{sec:comp}.

%to choose between the regularised estimation and rotation methods. 

%With the connection established above, 

%Let $\hat{\ttt}_R$ denote the oblique rotation solution under the CLC in \eqref{eq:11}, and let $\hat{\ttt}_\gamma$ denote the regularized solution with tuning parameter $\gamma$ as in \eqref{eq:4}. Suppose the solution path $\hat{\ttt}_\gamma$ converges when $\gamma \rightarrow 0 + $. Then 
%
%$
%\hat{\ttt}_\gamma\rightarrow\hat{\ttt}_R,  
%$ 
%
%when $\gamma \rightarrow 0+$.

\section{Statistical Inference and Asymptotic Theory}\label{sec:inf}

\subsection{Estimation Consistency}

We establish the statistical consistency of the proposed estimator based on the $L^p$ rotation.  Suppose the true parameter set
that we aim to recover is 
$(\mathbf{\Lambda^*},\mathbf{\Phi^*}, \mathbf{\Omega^*})$, where the true loading matrix $\mathbf{\Lambda^*}$ is sparse.
To emphasise the dependence on the sample size, we attach the sample size $N$ as a subscript to the initial estimator in the first step of the rotation method; that is, 
$\hat{\boldsymbol{\theta}}_N = (\hat{\mathbf{A}}_N, \mathbf{I},\hat{\boldsymbol{\Omega}}_N)$. We require the initial estimator to be consistent, in the sense that 

\begin{enumerate}
\item[C1.]   $\mathbf{\hat{A}}_N  \mathbf{\hat{A}}_N' \overset{pr}{\to} \mathbf{\Lambda^*\Phi ^*}\mathbf{\Lambda^* }'$ and $  \hat{\boldsymbol{\Omega}}_N \overset{pr}{\to} \boldsymbol{\Omega}^*$, where the notation ``$\overset{pr}{\to}$" denotes convergence in probability.  %\yc{(I changed the notation to $pr$, to avoid confusion with the p in the loss function.)}
\end{enumerate}

This requirement easily holds when the linear factor model is correctly specified and the loss function $L(\boldsymbol{\Sigma(\theta)})$ is reasonable (e.g., the negative log-likelihood). In addition, we require that the EFA model is truly a $K$-dimensional model, in the sense that condition C2 holds. 

\begin{enumerate}
    \item[C2.]  $rank(\mathbf{\Lambda^*\Phi ^*}\mathbf{\Lambda^* }') =K$. 
\end{enumerate}

For \yx{the} $L^p$ rotation estimator to be consistent, for a specific value of $p \in (0, 1]$, we further require that the true loading matrix uniquely minimises the $L^p$ CLF, in the sense of condition C3 below.
\begin{enumerate}
\item[C3.]   

$( \boldsymbol\Lambda^*, {\boldsymbol\Phi}^*) \in \argmin_{\boldsymbol\Lambda, \boldsymbol\Phi} Q_p(\boldsymbol\Lambda)  \mbox{~such that~} \boldsymbol\Lambda \boldsymbol\Phi\boldsymbol\Lambda' = \boldsymbol\Lambda^* \boldsymbol\Phi^*\boldsymbol\Lambda^{*'}.$ In addition, for any other $( \boldsymbol\Lambda^\dagger, {\boldsymbol\Phi}^\dagger) \in \argmin_{\boldsymbol\Lambda, \boldsymbol\Phi} Q_p(\boldsymbol\Lambda) \mbox{~such that~} \boldsymbol\Lambda \boldsymbol\Phi\boldsymbol\Lambda' = \boldsymbol\Lambda^* \boldsymbol\Phi^*\boldsymbol\Lambda^{*'}$, there exist $\mathbf D \in \mathcal D_1$ and $\tilde{\mathbf D} \in \mathcal D_2$, such that
${\boldsymbol\Lambda}^\dagger \mathbf D\tilde{\mathbf D} = \boldsymbol\Lambda^*$ 
and $\tilde{\mathbf D}^{-1} \mathbf  D^{-1}{\boldsymbol\Phi}^\dagger (\mathbf D^{-1})'(\tilde{\mathbf  D}^{-1})' = \boldsymbol\Phi^*$. Recall that $\mathcal D_1$ and $\mathcal D_2$ are the sets of column permutation and sign flip transformations, respectively, which we gave in Remark~\ref{rmk:indet}.
\end{enumerate}

Condition C3 tends to hold when the true loading matrix contains many zeros, as the $L^p$ loss function is a good approximation to the $L^0$ function that counts the number of non-zero elements. In particular, according to Proposition~\ref{prop:simple},  condition C3 is guaranteed to hold when $\boldsymbol\Lambda^*$ has a perfect simple structure, i.e., if it has at most one non-zero loading in each row. As we discussed in Section~\ref{subsec:propose}, this condition is more likely to hold for a smaller value of $p$, when there are cross-loadings.  
Conditions C1 through C3 guarantee the estimation consistency of the $L^p$ rotation estimator, up to column permutation and sign flips. We summarise this result in Theorem~\ref{thm:consistency} below. 

\begin{theorem}\label{thm:consistency} 
Suppose that for a given $p\in (0, 1]$ conditions C1 through C3 hold. Then there exist $\mathbf D_{N} \in \mathcal D_1$ and 
$\tilde{\mathbf D}_{N} \in \mathcal D_2$, such that 
%Under condition $\mathbf{C 1-5}$, $\mathbf{\hat{T}}_N \overset{p}{\to} \mathcal{T}^*$ and  $ \mathbf{\hat{\Lambda}}_N = \mathbf{\hat{A}}_N\mathbf{\hat{T}'}_N^{-1}  \overset{p}{\to}  \mathcal{H}^*, N \rightarrow \infty$
$\hat{\boldsymbol\Lambda}_{N,p} \mathbf D_{N}\tilde{\mathbf D}_{N} \overset{pr}{\to} \boldsymbol\Lambda^*$ and $\tilde{\mathbf D}_{N}^{-1} \mathbf  D_{N}^{-1}\hat{\boldsymbol\Phi}_{N,p} (\mathbf D_{N}^{-1})'(\tilde{\mathbf  D}_{N}^{-1})' \overset{pr}{\to} \boldsymbol\Phi^*$, where
\begin{equation*}%\label{eq:rotation2}
  (\hat{\boldsymbol\Lambda}_{N,p}, \hat{\boldsymbol\Phi}_{N,p}) \in \argmin_{\boldsymbol\Lambda, \boldsymbol\Phi} Q_p(\boldsymbol\Lambda), \mbox{~such that~} \boldsymbol\Lambda \boldsymbol\Phi\boldsymbol\Lambda' = \hat{\mathbf A}_N\hat{\mathbf A}_N'.
\end{equation*}
\end{theorem}

\subsection{Model Selection} 

The interpretation of the factors relies on the sign pattern of the loading matrix, so that we can interpret each factor based on the associated manifest variables and their directions (positive or negative associations). 
Learning this sign pattern is a model selection problem. A regularised estimator may seem advantageous as it yields simultaneous parameter estimation and model selection. We note that, however, we can easily achieve model selection with a rotation method, using a Hard-Thresholding (HT) procedure. Similar HT procedures have been proven to be successful in the model selection for linear regression models
\citep{meinshausen2009lasso}.

More precisely, let $\boldsymbol\Gamma^* = \left(\mbox{sgn}(\lambda_{jk}^*)\right)_{J\times K}$ denote the true sign pattern of $\boldsymbol\Lambda^*$, where $\mbox{sgn}(x)$ returns the sign of a scalar satisfying that  
$$\mbox{sgn}(x) = \left\{\begin{array}{ccc}
    1 & \mbox{~if~} x > 0,  \\
    0 & \mbox{~if~} x = 0,  \\
    -1& \mbox{~if~} x < 0. 
\end{array}\right.$$
Given the $L^p$ rotation estimator $\hat{\boldsymbol\Lambda}_{N,p} = \left(\hat \lambda_{jk}^{(N,p)}\right)_{J\times K}$, 
the HT procedure estimates the pattern of $\boldsymbol\Gamma^*$ by 
$\hat{\boldsymbol\Gamma}_{N,p} = \left(\mbox{sgn}(\hat \lambda_{jk}^{(N,p)}) \times 1_{\{|\hat \lambda_{jk}^{(N,p)}|>c\}}\right)_{J\times K}$, where $c > 0$ is a pre-specified threshold.  If we choose the threshold $c$ properly, then $\hat{\boldsymbol\Gamma}_{N,p} $ consistently estimates 
$\boldsymbol\Gamma^*$. We state this result in Theorem~\ref{thm:selection} below. 

\begin{enumerate}
    \item[C4.] The threshold $c$ lies in the interval $(0, c_0)$, where 
    $c_0 = \min\{ |\lambda_{jk}^*|: \lambda_{jk}^*\neq 0\}$. 
\end{enumerate}

\begin{theorem}\label{thm:selection}
Suppose that for a given $p\in (0, 1]$ conditions C1 through C4 hold. Then there exist  $\mathbf D_{N} \in \mathcal D_1$ and $\tilde{\mathbf D}_{N} \in \mathcal D_2$, such that the probability
$P(\hat{\boldsymbol\Gamma}_{N,p}\mathbf D_{N}\tilde{\mathbf D}_{N} = \boldsymbol\Gamma^*)$
converges to 1 as the sample size $N$ goes to infinity. 
\end{theorem}

%determines the non-zero pattern of the loading matrix, by comparing the estimates with a threshold $c > 0$. 
In practice, the value of $c_0$ is unknown and thus cannot be used for choosing the threshold $c$. Instead, 
we choose  $c$ based on the Bayesian Information Criterion
\citep[BIC;][]{schwarz1978estimating}. We summarise the steps
of this procedure in Algorithm~\ref{alg:bic} below, where we simplify the notation for ease of exposition. 

\begin{algorithm}
\caption{Hard-thresholding for model selection based on $L^p$ rotation}\label{alg:bic}
\textbf{Input:} A sequence of candidate thresholds $\mathcal C$, observed data, and the rotated loading matrix $\hat{\boldsymbol\Lambda} = (\hat \lambda_{jk})_{J\times K}$ given by the $L^p$ CLF criterion. 

\medskip
For each value of $c \in \mathcal C$, we perform the following two steps: \begin{enumerate}
    \item[] \textbf{Step 1:} Obtain the corresponding selected loading structure $\hat{\boldsymbol\Gamma}_c = \left( \mbox{sgn}(\hat\lambda_{jk})\times 1_{\{|\hat \lambda_{jk}|>c\}}\right)_{J\times K}$.
 
\item[] \textbf{Step 2:} Fit a Confirmatory Factor Analysis (CFA) model based on $\hat{\boldsymbol\Gamma}_c$ using the maximum likelihood estimator, in which the $(i,j)$th loading parameter satisfies the sign constraint implied by the corresponding entry of $\hat{\boldsymbol\Gamma}_c$.   Calculate the BIC value for this CFA model, denoted by $\mbox{BIC}_c$. 
\end{enumerate}

Obtain 
$\hat c = \argmin_{c\in \mathcal C} \mbox{BIC}_c.$

\medskip
\textbf{Output:} The selected sign pattern $\hat{\boldsymbol\Gamma}_{\hat c}$. 
\end{algorithm}

When the candidate values of $c$ are chosen properly (i.e., $\mathcal C$ includes values that are below $c_0$), then Theorem~\ref{thm:selection} implies that with probability tending to one, the true model will be in the candidate models. Together with the consistency of BIC for parametric models \citep{shao1997asymptotic,vrieze2012model}, the true non-zero pattern can be consistently recovered. \yc{We remark that it may not be a good idea to manually select $c$ or use some default thresholds. Unless there is very good substantive knowledge about the latent structure, it is very likely to under- or over-select $c$, leading to high false-positive and false-negative errors. 
Even with the proposed procedure, the selection consistency is only guaranteed when the sample size goes to infinity. For a finite sample, the 
false-positive and false-negative errors likely exist and thus we should look at the selected model with caution. Furthermore, we note that the BIC is not the only information criterion that leads to model selection consistency \citep{nishii1984asymptotic}, but it is probably the most commonly used information criterion with consistency guarantee. Another commonly used information criterion is the Akaike Information Criterion (AIC) which tends to over-select and thus does not guarantee model selection consistency \citep{shao1997asymptotic}. 
 
}

%\yx{We point out that that the proposed model selection procedure is based on a large-sample argument. However, in constrast to e.g. the Akaike information criterion, the BIC penalizes more and consistently selects the model. If we instead manually select $c$ but set it too small while the noise level is large, we would consistently select the wrong model. Since we do not know the noise level, we use the BIC as it consistently selects the model.}

\subsection{Confidence Intervals}

Often, we are not only interested in the point estimate of the underlying sparse loading matrix, but also in quantifying its uncertainty. We typically achieve uncertainty quantification by constructing confidence intervals for the loadings of the rotated solution. Traditionally, we can do this by establishing the asymptotic normality of the rotated loading matrix using the delta method, which involves calculating the partial 
derivatives of a rotation algorithm using implicit differentiation \citep{jennrich1973standard}. Unfortunately, this procedure is no longer suitable if the true loading matrix is sparse and the loss function is not differentiable with respect to the zero loadings. 

Motivated by a simple but nevertheless well-performing post-selection inference procedure in regression analysis \citep{zhao2021defense}, we propose a procedure for constructing confidence intervals for individual loading parameters of the rotated solution. More precisely, this procedure runs a loop over all the manifest variables, $j=1, ..., J$. Each time, the procedure obtains the confidence intervals for the loading parameters of manifest variable $j$ by fitting a CFA model whose loading structure is determined by the selected sign pattern  of the \yc{remaining} $J-1$ manifest variables. More precisely, the loading parameters of the CFA model satisfy the sign constraints imposed by  the selected sign pattern $\hat{\boldsymbol\Gamma}_{\hat c}$ from Algorithm~\ref{alg:bic}, for all the items except for $j$. We impose no constraint on the loading parameters of item $j$. We obtain confidence intervals for the loading parameters of item $j$ based on the asymptotic normality of the estimator for this CFA model. 
%of the rest of the manifest variables is determined by the selected sign pattern $\hat{\boldsymbol\Gamma}$ and the loading parameters of manifest variable $j$ are freely estimated. 
%whose loading structure is determined by the 
We summarise this procedure in Algorithm~\ref{alg:CI} below.

\begin{algorithm} 
\caption{Post-selection confidence intervals}\label{alg:CI}
\textbf{Input:}  The selected sign pattern $\hat{\boldsymbol\Gamma} = (\hat \gamma_{jk})_{J\times K}$, observed data, and significance level $\alpha \in (0,1)$. 

\medskip

For each manifest variable $s = 1, ..., J$, we perform the following two steps: 
\begin{enumerate}
    \item[] \textbf{Step 1:} Obtain a CFA model whose loadings $\lambda_{jk}$ satisfy the constraints that 
    $\mbox{sgn}(\lambda_{jk}) = \hat \gamma_{jk}$ for all $j \neq s$ and for all $k$. 
 
\item[] \textbf{Step 2:}  Fit the CFA model and obtain the $(1-\alpha)$-confidence intervals for parameters $\lambda_{s1}$, ..., $\lambda_{sK}$ using a standard inference procedure for CFA (e.g., based on the maximum likelihood estimator). We denote these confidence intervals by $(l_{sk}, u_{sk})$. If the CFA model in \textbf{Step 1} is not identifiable, we let the confidence intervals \yc{be} $(-\infty, \infty)$. 
\end{enumerate}
\textbf{Output:} Confidence intervals $(l_{sk}, u_{sk})$, $s = 1, ..., J, k=1, ..., K$. 
\end{algorithm}

In what follows, we establish the consistency of confidence intervals given by Algorithm~\ref{alg:CI}. To emphasise that the statistics in Algorithm~\ref{alg:CI} depend on the sample size $N$, we attach $N$ as a subscript or superscript when describing this consistency result. We require the following conditions: 

\begin{enumerate}
    \item[C5.] The selected sign pattern $\hat{\boldsymbol\Gamma}_N$ is consistent. That is, there exist  $\mathbf D_{N} \in \mathcal D_1$ and $\tilde{\mathbf D}_{N} \in \mathcal D_2$, such that the probability
$P(\hat{\boldsymbol\Gamma}_{N,p}\mathbf D_{N}\tilde{\mathbf D}_{N} = \boldsymbol\Gamma^*)$
converges to 1 as the sample size $N$ goes to infinity.
\end{enumerate}
Thanks to the consistency of BIC selection, and when we have chosen the candidate thresholds properly, condition C5 holds if $\hat{\boldsymbol\Gamma}_N$ is obtained by Algorithm \ref{alg:bic}. 

\begin{enumerate}
    \item[C6.]  For each manifest variable $s = 1, ..., J$, the CFA model whose loading parameters satisfy
    $\mbox{sgn}(\lambda_{jk}) =  \mbox{sgn}(\lambda_{jk}^*)$ for all $j \neq s$ is identifiable, and using the same procedure in Step 2 of Algorithm~\ref{alg:CI} leads to consistent confidence intervals for $\lambda_{s1}$, ..., $\lambda_{sK}$.  That is, let $(l_{sk}^{*(N)}, u_{sk}^{*(N)})$ be the resulting confidence interval for $\lambda_{sk}$, then
    $P(\lambda_{sk}^* \in (l_{sk}^{*(N)}, u_{sk}^{*(N)}))$ converges to $1-\alpha$, as the sample size $N$ goes to infinity.  
\end{enumerate}
Note that C6 is a condition imposed on the sign pattern of the true loading matrix. It essentially requires that the factors can be identified by the sign pattern of any $(J-1)$-subset of the manifest variables. Given an identified CFA model, we can easily construct the consistent confidence intervals based on the asymptotic normality of any reasonable estimator of the CFA model, e.g., the maximum likelihood estimator. 
Under conditions C5 and C6, the following theorem holds.

\begin{theorem}\label{thm:ci}
Suppose that conditions C5 and C6 hold for the selected sign pattern $\hat{\boldsymbol\Gamma}_N$ and the true model, where  $\mathbf D_{N} \in \mathcal D_1$ and 
$\tilde{\mathbf D}_{N} \in \mathcal D_2$ are from condition C5. Suppose we input $\hat{\boldsymbol\Gamma}_N$, observed data from the true model, and significance level $\alpha$ into the true model, and obtain output 
$(l_{sk}^{(N)}, u_{sk}^{(N)})$, $s = 1, ..., J, k=1, ..., K$. 
Then we have 
$P(\lambda_{sk}^{*(N)} \in (l_{sk}^{(N)}, u_{sk}^{(N)}))$ 
converges to $1-\alpha$, for all  $s = 1, ..., J, k=1, ..., K$, 
where $\lambda_{sk}^{*(N)}$ are entries of $\boldsymbol\Lambda^{*(N)} = \boldsymbol\Lambda^* \tilde{\mathbf D}_{N}^{-1}\mathbf D_{N}^{-1}$. Note that 
$\boldsymbol\Lambda^{*(N)}$ is equivalent to $\boldsymbol\Lambda^{*}$ up to column permutation and sign flips. 

We remark that under the conditions of Theorem  \ref{thm:ci}, all the CFA models fitted in Step 2 of Algorithm \ref{alg:CI} should be identifiable for sufficiently large $N$. However, in practice, it may happen that some CFA models are not identifiable, either due to the sample size not being large enough or the regularity conditions C5 or C6 not holding. In such cases, we set the corresponding confidence intervals  to be $(-\infty, \infty)$ as a conservative choice. 

\end{theorem}

\section{Computation}\label{sec:comp}

\subsection{Proposed IRGP Algorithm}
We now discuss the computation for the proposed rotation. 
Recall that we aim to solve the optimisation problem 
$$\mathbf{\hat{T}} \in \argmin_{ \mathbf{T} \in \mathcal{M}} Q_p(\hat{\mathbf{A}} \mathbf{T'}^{-1}),$$
where $Q_p$ is the $L^p$ CLF defined in \eqref{eq:CLC_Lp}. Note that this objective function is not differentiable when $\hat{\mathbf{A}} \mathbf{T'}^{-1}$ has zero elements, as the $L^p$ function is not smooth at zero. Consequently, standard numerical solvers fail, especially when the true solution is approximately sparse. To solve this optimisation problem, we develop an IRGP algorithm that combines the iteratively reweighted least square algorithm \citep{ba2013convergence,daubechies2010iteratively} and the gradient projection algorithm \citep{jennrich2002simple}. 

Similar to \citet{jennrich2006rotation}, the IRGP algorithm also solves a smooth approximation to the objective function $Q_p(\hat{\mathbf{A}} \mathbf{T'}^{-1})$. That is, we introduce a sufficiently small constant $\epsilon > 0$, and approximate the objective function by $Q_{p,\epsilon}(\hat{\mathbf{A}} \mathbf{T'}^{-1})$,
where 
$$Q_{p,\epsilon}(\boldsymbol\Lambda) = \sum_{j=1}^J\sum_{k=1}^K (\epsilon^2 + \lambda_{jk}^2)^{\frac{p}{2}}.$$ 
As we discuss in the sequel, the $\epsilon$ is introduced to make the computation more robust. The IRGP algorithm alternates between two steps -- (1) function approximation step and (2) Projected Gradient Descent (PGD) step. More precisely, let 
$T_t$ be the parameter value at the $t$th iteration.

The function approximation step involves approximating the objective function by 
\begin{equation}\label{eq:appr}
G_t(\mathbf T) = \sum_{j=1}^J\sum_{k=1}^K w_{jk}^{(t)} \left((\hat{\mathbf{A}} \mathbf{T'}^{-1})_{jk}\right)^2,
\end{equation}
where the weights $w_{jk}^{(t)}$ are given by 
\yc{$$w_{jk}^{(t)} = \frac{1}{((\hat{\mathbf{A}} (\mathbf{T}_t')^{-1})_{jk}^2 + \epsilon^2 )^{1-p/2}}.$$}
Here  $\epsilon > 0$ is a pre-specified parameter that is chosen to be sufficiently small. We provide some remarks about  this approximation. First, the small tuning parameter is chosen to stabilise the algorithm when certain  $\hat{\mathbf{A}} (\mathbf{T}_t')^{-1})_{jk}$s are close to zero. Without $\epsilon$, the weight 
$w_{jk}^{(t)}$ can become very large, resulting in an unstable PGD step. Second, supposing that $(\hat{\mathbf{A}} (\mathbf{T}_t')^{-1})_{jk} \neq 0$ for all $j$ and $k$, then $G_t(\mathbf{T}_t) \approx Q_p(\hat{\mathbf{A}} (\mathbf{T}'_t)^{-1})$ when $\epsilon$ is sufficiently small, i.e., the function approximation and the objective function value are close to each other at the current 
parameter value. Lastly, this approximation is similar to the E-step of the Expectation-Maximisation algorithm \citep{dempster1977maximum}; see \citet{ba2013convergence} for this correspondence. 

The PGD step involves updating the parameter value based on the $G_t(\mathbf T)$ via projected gradient descent. 
This step is similar to the update in each iteration of the gradient projection algorithm for oblique rotations \citep{jennrich2002simple}. 
We can perform PGD on $G_t(\mathbf T)$, as this function approximation is smooth in $\mathbf T$. More precisely, we define a projection operator as 
\begin{equation}\label{eq:proj}
    \mbox{Proj}(\mathbf T) = \mathbf T (\mbox{diag}(\mathbf T' \mathbf T))^{-\frac{1}{2}},
\end{equation}
where $(\mbox{diag}(\mathbf T' \mathbf T))^{-\frac{1}{2}}$ is a diagonal matrix whose $i$th diagonal entry is given by 
$1/\sqrt{(\mathbf T' \mathbf T)_{ii}}$. This operator projects any invertible matrix into the space of oblique rotation matrices $\mathcal M$ \yc{as defined in \eqref{eq:13}}. The PGD update is given by 
\begin{equation}\label{eq:PGD}
\mathbf T_{t+1} = \mbox{Proj}(\mathbf T_t - \alpha \grad G_t(\mathbf T)), 
\end{equation}
where $\alpha > 0$ is a step size chosen by line search
and $\grad G_t(\mathbf T)$ is a $K\times K$ matrix whose $(i,j)$th entry is the partial derivative of $G_t(\mathbf T)$ with respect to the $(i,j)$th entry of $\mathbf T$. We summarise the IRGP algorithm below.

\begin{algorithm}
\caption{IRGP algorithm for $L^p$ rotation}\label{alg:cap}
\textbf{Input:} The initial loading matrix estimate $\hat{\mathbf{A}}$, parameter $\epsilon >0$, 
and an initial value $\mathbf T_0$. 

\medskip

For iterations $t = 0, 1, 2, ...$, we iterate between the following two steps: 

\begin{enumerate}
\item[] \textbf{Step 1:} Construct $G_t(\mathbf T)$ using equation \eqref{eq:appr}. 
\item[] \textbf{Step 2:} Obtain $\mathbf T_{t+1}$  using equation \eqref{eq:PGD}, where the step size $\alpha$ is chosen by line search. 
\end{enumerate}

Stop when the convergence criterion is met. Let $t_{max}$ be the final iteration number. 

\medskip
\textbf{Output:} $\mathbf{T}_{t_{max}}$.  
\end{algorithm}

Under reasonable regularity conditions \citep{ba2013convergence}, every limit point of $\{\mathbf T_t\}_{t=1}^\infty$ will be a stationary point of the approximated objective function 
$Q_{p,\epsilon}(\hat{\mathbf{A}} \mathbf{T'}^{-1})$.
In addition, the algorithm has local linear convergence 
when $p = 1$ and super-linear convergence when $0<p<1$. 

We remark on the choice of initial value $\mathbf T_0$ when $0< p <1$. As discussed previously in Section~\ref{subsec:propose}, when $0< p <1$, the objective function $Q_p(\hat{\mathbf{A}} \mathbf{T'}^{-1})$ is highly non-convex and thus may contain many stationary points. To avoid the algorithm getting stuck at a local optimum, the choice of $\mathbf T_0$ is important. 
When solving the optimisation for a smaller value of $p$, we recommend using the solution from a larger value of $p$ as the starting point (e.g., $p=1$).

%the optimisation problem 

\subsection{Comparison with Regularised Estimation} 

To compare the computation of the proposed rotation method and that of regularised estimation, we also describe a proximal gradient algorithm for the $L^1$ regularised estimator.
The proximal algorithm is a state-of-the-art algorithm for solving nonsmooth optimisation problems \citep{parikh2014proximal}. We can view it as a generalisation of projected gradient descent. As we will discuss below, each iteration of the algorithm can be computed easily. In principle, we can also apply the proximal algorithm to the $L_p$ regularised estimator, for $0 < p <1$. However, it is computationally much more costly than the case when $p=1$, and thus, will not be discussed here. 

The $L^1$ regularised estimator, also referred to as the LASSO estimator, solves the following optimisation problem:
$$\min_{\ttt}~~ L(\boldsymbol{\Sigma}(\boldsymbol\theta)) + \gamma  \sum_{j=1}^J\sum_{k=1}^K |\lambda_{jk}|.$$
To apply the proximal gradient algorithm, we reparameterise the covariance matrix $\boldsymbol\Phi$ by 
$\mathbf T' \mathbf T$, \yc{where we let $\mathbf T$ be an upper triangular matrix to ensure its identifiability}. We also reparameterise the diagonal entries of the diagonal matrix $\boldsymbol\Omega$ by $\mathbf v = (v_{1}, ..., v_{J})$, where $v_i = \log(\omega_{ii})$. With slight abuse of notation, we can write the optimisation problem as 
$$\min_{\boldsymbol\Lambda, \mathbf T, \mathbf v}~~ L(\boldsymbol{\Sigma}(\boldsymbol\Lambda, \mathbf T, \mathbf v)) + \gamma  \sum_{j=1}^J\sum_{k=1}^K |\lambda_{jk}|.$$

We define a proximal operator for the loading matrix as 
\begin{equation}\label{eq:prox}
\mbox{Prox}_{\alpha, \gamma}(\tilde{\boldsymbol \Lambda}_t) = \argmin_{\boldsymbol\Lambda} ~~\frac{1}{2} \sum_{j=1}^J\sum_{k=1}^K (\lambda_{jk} - \tilde\lambda_{jk}^{(t)})^2 + \alpha\gamma \sum_{j=1}^J\sum_{k=1}^K |\lambda_{jk}|,
\end{equation}
where $\alpha > 0$ will be a step size  and $\tilde{\boldsymbol \Lambda}_t = (\tilde\lambda_{jk}^{(t)})_{J\times K}$
will be the value of $\boldsymbol \Lambda$ from the previous step in the proximal gradient algorithm. Note that  \eqref{eq:prox} has a closed-form solution given by soft-thresholding \citep{parikh2014proximal} that we can easily compute. We summarise the proximal gradient algorithm in Algorithm
\ref{alg:prox} below.

\begin{algorithm}
\caption{Proximal gradient algorithm for $L^1$ regularised estimation.}\label{alg:prox}
\textbf{Input:} The initial values $\boldsymbol \Lambda_0$, $\mathbf T_0$, and $\mathbf v_0$. 

\medskip

For iterations $t = 0, 1, 2, ...$, we iterate between the following two steps: 

\begin{enumerate}
\item[] \textbf{Step 1:}  Calculate the gradients of $L(\boldsymbol{\Sigma}(\boldsymbol\Lambda, \mathbf T, \mathbf v))$  with respect to $\boldsymbol\Lambda$, $\mathbf T$, and $\mathbf v$, respectively, at $(\boldsymbol \Lambda_t, \mathbf T_t, \mathbf v_t)$. Denote these gradients by $\grad L_{t,\boldsymbol\Lambda}$
$\grad L_{t,\mathbf T}$, and $\grad L_{t,\mathbf v}$.

\item[] \textbf{Step 2:} Update the parameters by 
$$\boldsymbol \Lambda_{t+1} = \mbox{Prox}_{\alpha, \gamma}(\boldsymbol \Lambda_t - \alpha \grad L_{t,\boldsymbol\Lambda}),$$ 
$$\mathbf T_{t+1} = \mbox{Proj}(\mathbf T_t - \alpha \grad L_{t,\mathbf T}),$$
and 
$$\mathbf v_{t+1} = \mathbf v_t - \alpha \grad L_{t,\mathbf v}.$$

Recall that the operator $\mbox{Proj}(\cdot)$ is defined in \eqref{eq:proj}, and $\alpha$ is a step size chosen by line search. 

\end{enumerate}

Stop when the convergence criterion is met. Let $t_{max}$ be the final iteration number. 

\medskip
\textbf{Output:} $(\boldsymbol \Lambda_{t_{max}}, \mathbf T_{t_{max}}, \mathbf v_{t_{max}})$. %$\mathbf{T}_{t_{max}}$.  
\end{algorithm}
Under suitable conditions, this proximal gradient algorithm converges to stationary points of the objective function and has a local linear convergence rate \citep{karimi2016linear}. 
We notice that when $p=1$, Algorithms~\ref{alg:cap} and \ref{alg:prox} have similar convergence properties.  However, their per-iteration computational complexities 
are different. In particular,  Algorithm \ref{alg:prox}
involves parameters $\boldsymbol\Lambda$ and $\mathbf v$, which substantially increases its computational complexity. In fact, the per-iteration complexity for Algorithm~\ref{alg:cap} is $O(K^3+K^2J)$, while that for Algorithm~\ref{alg:prox} is $O(J^3+J^2K+K^2J+ {K^3})$. The difference can be substantial when $J$ is much larger than $K$. \yc{We give the derivation of these computational complexities in the supplementary material.} 

\section{Simulation Study}\label{sec:sim}

\subsection{Study I}

In this study, we evaluate the performance of $L^{0.5}$ and $L^1$ rotations and compare them with some traditional rotation methods and $L^1$-regularised estimation. We consider several traditional oblique rotation methods, including the oblimin, quartmin, simplimax, geomin, and promax methods. These methods have been considered in the simulation studies in \citet{jennrich2006rotation}. 
They are implemented using the \texttt{GPArotation} package \citep{bernaards2005gradient} in R. 

\paragraph{Settings.} 
We consider two simulation settings, one with $J=15$ manifest variables and $K=3$ factors, and the other with $J=30$ and $K=5$.  The first setting has nine manifest variables each \yx{loading} on a single factor (three variables for each factor), and six manifest variables each \yx{loading} on two factors. The second setting has 15 manifest variables each \yx{loading} on a single factor (three variables for each factor), 10 manifest variables each \yx{loading} on two factors, and 5 manifest variables each \yx{loading} on three factors. We give the true model parameters in the supplementary material.  
By numerical evaluations, the true loading matrices satisfy condition C3 for both $L^{0.5}$ and $L^1$ criteria. Under each setting, we consider three sample sizes, including $N = 400, 800$, and 1,600. 
For each setting and each sample size, we run $B = 500$ independent replications. 

%~\yc{Describe the simulation settings. and the number of replications we use. The true model parameters should be given in the supplement. Here, only describe the setting of the parameters.  You point out that via numerical evaluations, the true loading matrix satisfies condition C3.}

\paragraph{Evaluation criteria.} We evaluate the proposed method from three aspects. First, we compare all estimators in terms of accuracy of point estimation. Second, we compare the proposed method and the $L^1$ regularised estimator in terms of their model selection accuracy. Finally, we examine the coverage rate of the proposed method for constructing confidence intervals.

When evaluating the performance of different estimators,
we take into account the indeterminacy due to column permutations and sign flips. Let $\tilde{\mathbf{\Lambda}}^{(b)}$ be the loading matrix estimate given by a rotation or regularised estimation method in the $b$th replication. We then find 
$$\hat{\mathbf{\Lambda}}^{(b)} = \argmin_{\mathbf{\Lambda}}\{\Vert\mathbf{\Lambda} - \mathbf{\Lambda}^*\Vert^2: \mathbf{\Lambda} =\tilde{\mathbf{\Lambda}}^{(b)}\mathbf D \tilde{\mathbf D}, \mathbf D \in \mathcal D_1,\tilde{\mathbf D} \in \mathcal D_2 \},$$
which is the one closest to the true loading matrix $\mathbf{\Lambda}^*$ among all the loading matrices that are equivalent to $\tilde{\mathbf{\Lambda}}^{(b)}$. Our evaluation criteria are constructed based on 
$\hat{\mathbf{\Lambda}}^{(b)}$:

%We list the corresponding evaluation criteria below:
 %\yc{For each one, describe the criteria (i.e., losses) that we calculate to compare methods. Use some NICE notations (some current ones are not elegant!).}
\begin{enumerate}

\item The accuracy of point estimation is estimated by the mean squared error (MSE):
%
%$$ 
%\text{MSE} = \sum (\hat{\lambda}^{(b)}_{ij} - \lambda_{ij}^*)^2 / $JK, 
%$$
$$ 
\text{MSE} =   \frac{||\hat{\mathbf{\Lambda}}^{(b)}  - \mathbf{\Lambda}^*||_F^2}{JK}, 
$$
where $\hat{\mathbf{\Lambda}}^{(b)}$
%\{ \hat{\lambda}^{(b)}_{ij} \}_{J \times K}
%
is obtained by a certain rotation or regularisation method in the $b$-th replication. %and 
% \{ \lambda^*_{ij} \}_{J \times K} = 
%$
%\mathbf{\Lambda}^*
%$ 
%
%is the true loading matrix. %\yx{how did you take into account the permutations and sign flips? The same for the following criteria?}

\item The model selection accuracy is assessed using the area under the curve (AUC) from the corresponding receiver operating characteristic (ROC) curve. For each threshold $c$, we compute the average true positive rate ($\widebar{\text{TPR}}_c$), which is the proportion of successfully identified non-zero elements in the true loading matrix:
\begin{equation}\label{eq:TPR}
 \widebar{\text{TPR}}_c = \frac{1}{B} \sum_{b=1}^B\frac{\sum_{j,k}\mathbbm{1}_{ \{\hat{\lambda}^{(b,c)}_{jk} \neq 0, \lambda_{jk}^* \neq 0 \}} }{\sum_{j,k} \mathbbm{1}_{\{\lambda_{jk}^* \neq 0\}}},
\end{equation}
where  
$ \{ \hat{\lambda}^{(b, c)}_{jk} \}_{J \times K}=
 \hat{\mathbf{\Lambda}}^{(b,c)} 
$ 
is the estimated loading matrix in the $b$-th replication from a CFA model based on $\hat{\boldsymbol\Gamma}_c$ using the maximum likelihood estimator.  Similarly, we calculate the average true negative rate ($\widebar{\text{TNR}}_c$), which is the success rate of identifying zero elements:
\begin{equation}\label{eq:TNR}
 \widebar{\text{TNR}}_c = \frac{1}{B} \sum_{b=1}^B \frac{\sum_{j,k}\mathbbm{1}_{ \{\hat{\lambda}^{(b,c)}_{jk} = 0, \lambda_{jk}^* = 0 \}} }{\sum_{j,k} \mathbbm{1}_{\{\lambda_{jk}^* = 0\}}}.
\end{equation}
The AUC is consequently calculated by plotting $\widebar{\text{TPR}}_c$ against $1 - \widebar{\text{TNR}}_c$ by varying the threshold $c$. We also use the overall selection accuracy, i.e., the true selection rate ($\text{TR}$), to evaluate the model selection procedure described in Algorithm~\ref{alg:bic}. The TR is calculated as
$$
\text{TR} = \frac{1}{B} \sum_{b=1}^B\frac{\sum_{j,k}\mathbbm{1}_{ \{\hat{\lambda}^{(b,\hat{c})}_{jk} \neq 0, \lambda_{jk}^* \neq 0 \}} + \sum_{j,k}\mathbbm{1}_{ \{\hat{\lambda}^{(b,\hat{c})}_{jk} = 0, \lambda_{jk}^* = 0 \}}}{JK},
$$
where $\hat{c}$ is the BIC selected threshold value from Algorithm~\ref{alg:bic}. Correspondingly, we calculate the $\text{TPR}$ and $\text{TNR}$ of the selected model as 
$$\text{TPR} = \widebar{\text{TPR}}_{\hat c} ~\mbox{and}~ \text{TNR} = \widebar{\text{TNR}}_{\hat c}.$$
\item The entry-wise 95\% confidence interval coverage rate (ECIC) is calculated to evaluate the performance of our post-selection confidence interval procedure in Algorithm~\ref{alg:CI}. For each entry of the loading matrix, the empirical probability of the true loading falling within the estimated confidence interval is calculated as 
$$
\text{ECIC}_{jk}=\frac{\sum_{b=1}^B \mathbbm{1}_{ \{ \lambda_{jk}^{*(N)} \in (l_{jk}^{(N)}, u_{jk}^{(N)})\} }}{B}.
$$
\end{enumerate}

\paragraph{Results on point estimation.} %~\yc{Present and discuss the results.}

In Table \ref{tab:MSE}, we present the MSE of the estimated loading matrix, for both simulation settings and $N \in \{400, 800, 1,600\}$. In the first five rows we show the results based on traditional oblique rotation criteria, followed by the results of the proposed $L^p$ loss function for two choices of $p$, and finally those of the LASSO estimator for five choices of $\gamma$. For both settings and all sample sizes, geomin performed the best among the traditional rotation methods. 
%The resulting MSEs did furthermore not differ between the settings.
The geomin results were very similar to those of $L^p$ rotation and the LASSO estimator with sufficiently small tuning parameter $\gamma$. For the LASSO estimator, the MSE increased as $\gamma$ increased. For $L^p$ rotation, we observed only very small differences between $p = 0.5$ and $p = 1$. In addition, their
MSEs were close to those of the LASSO estimator with $\gamma = 0.01$ and $\gamma = 0.05$.

\begin{table}[]
\centering
\caption{MSE obtained by using different rotation criteria under various settings, Study I.}
\label{tab:MSE}
\begin{tabular}{lcccccc}
\hline
 & \multicolumn{3}{c}{$15 \times 3$}               & \multicolumn{3}{c}{$30\times 5$} \\ \hline
                                            & $N=400$ & $N=800$ & $N=1,600$                   & $N=400$   & $N=800$  & $N=1,600$  \\ \hline
\multicolumn{1}{l|}{Oblimin}                & 0.012   & 0.007   & \multicolumn{1}{c|}{0.004} & 0.012     & 0.008    & 0.006     \\
\multicolumn{1}{l|}{GeominQ}                & 0.010   & 0.005   & \multicolumn{1}{c|}{0.002} & 0.010     & 0.005    & 0.002     \\
\multicolumn{1}{l|}{Promax}                 & 0.013   & 0.007   & \multicolumn{1}{c|}{0.005} & 0.014     & 0.009    & 0.007     \\

\multicolumn{1}{l|}{$L^{0.5}$ rotation}     & 0.011   & 0.005   & \multicolumn{1}{c|}{0.003} & 0.009     & 0.005    & 0.002     \\
\multicolumn{1}{l|}{$L^{1}$ rotation}       & 0.010   & 0.005   & \multicolumn{1}{c|}{0.003} & 0.010     & 0.004    & 0.002     \\
\multicolumn{1}{l|}{LASSO, $\gamma = 0.01$} & 0.009   & 0.004   & \multicolumn{1}{c|}{0.002} & 0.008     & 0.003    & 0.002     \\
\multicolumn{1}{l|}{LASSO, $\gamma = 0.05$} & 0.009   & 0.006   & \multicolumn{1}{c|}{0.005} & 0.007     & 0.005    & 0.004     \\
\multicolumn{1}{l|}{LASSO, $\gamma = 0.1$}  & 0.017   & 0.015   & \multicolumn{1}{c|}{0.014} & 0.012     & 0.011    & 0.010     \\
\multicolumn{1}{l|}{LASSO, $\gamma = 0.2$}  & 0.079   & 0.076   & \multicolumn{1}{c|}{0.074} & 0.038     & 0.034    & 0.032     \\
\multicolumn{1}{l|}{LASSO, $\gamma = 0.5$}  & 0.244   & 0.244   & \multicolumn{1}{c|}{0.244} & 0.144     & 0.149    & 0.150     \\ \hline
\end{tabular}
\end{table}
\begin{figure}[t]  
\centering
     \begin{subfigure}{.49\textwidth} 
         \centering         \includegraphics[width=\linewidth]{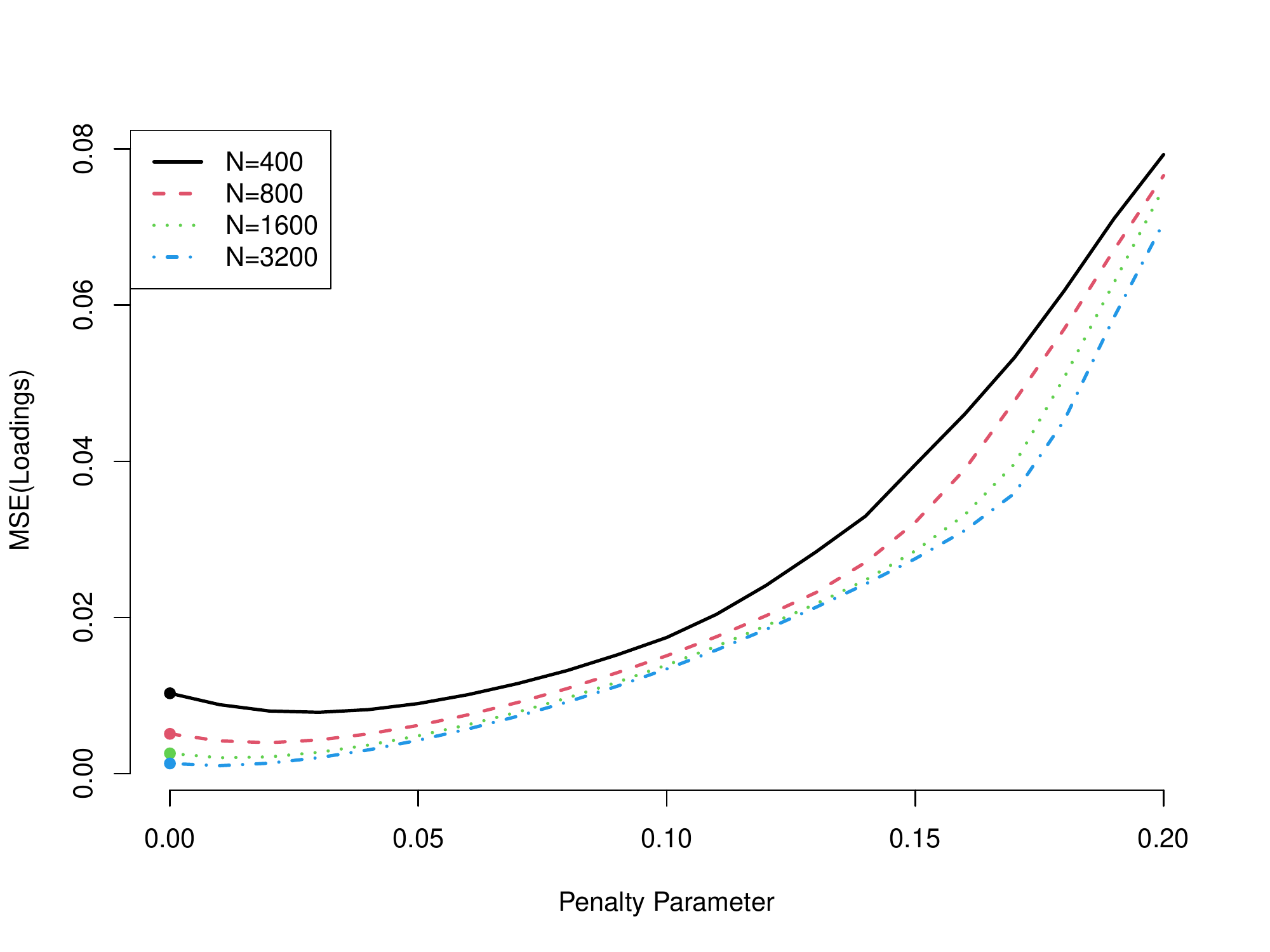}
         \caption{}
         %\label{fig:p=0.5}
     \end{subfigure}
     \hfill
     \begin{subfigure}{.49\textwidth}
         \centering
         \includegraphics[width=\linewidth]{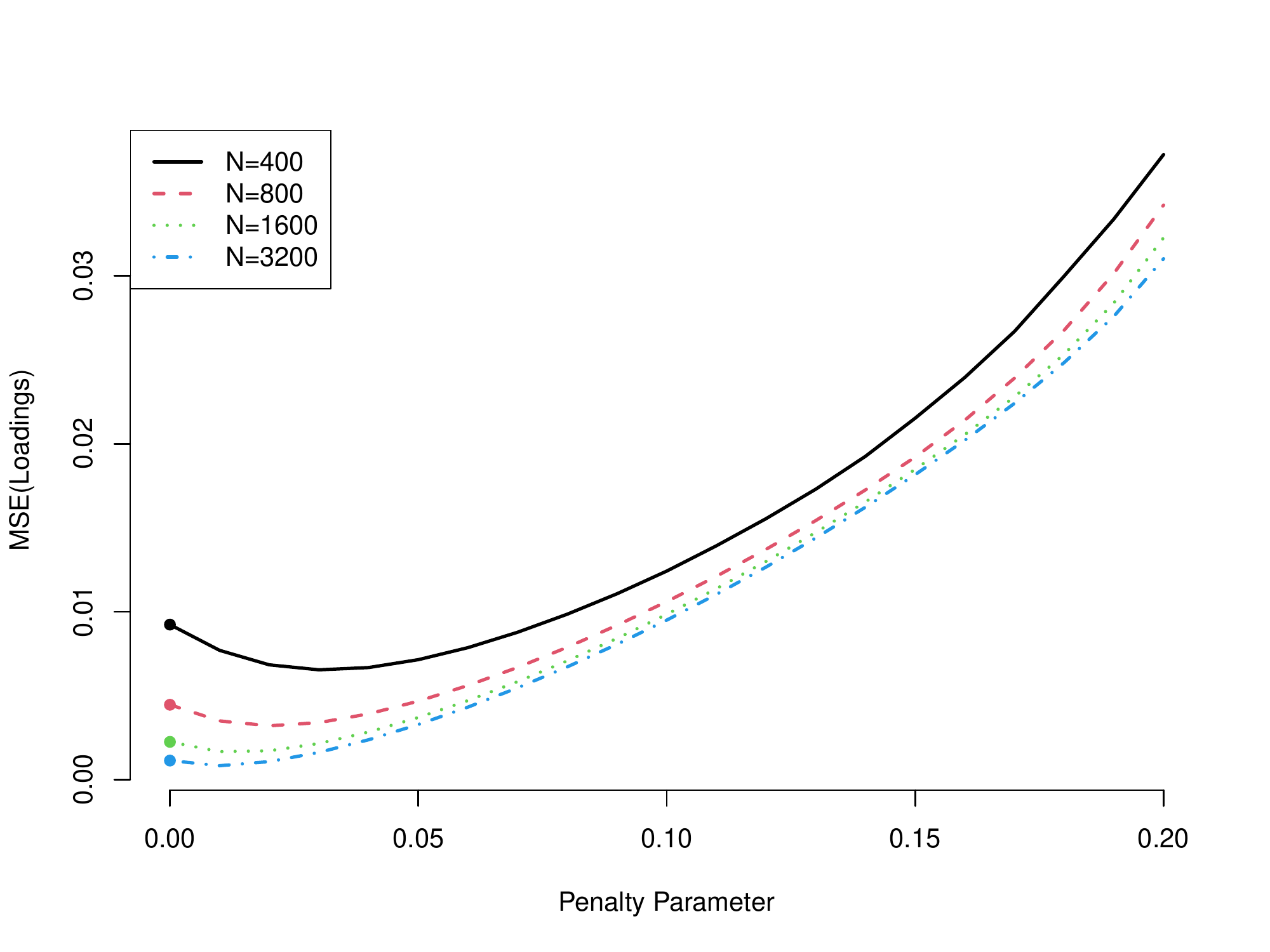}
         \caption{}
         %\label{fig:p=1}
     \end{subfigure}
        \caption{\yc{The MSE (for loadings) as a function of the tuning parameter $\gamma$ in the LASSO-regularised estimator. Panel (a): $15 \times 3$ settings. Panel (b): $30 \times 5 $ settings.   The dots at $\gamma = 0$ correspond to 
        the $L^1$ rotation solutions. }}
        \label{fig:path}
\end{figure}
\paragraph{Results on model selection.} 
In Table \ref{tab:modacc}, we present the AUC, TR, TPR, and TNR for the $L^p$ rotations and the LASSO estimator with different tuning parameters. 
For both scenarios and all sample sizes, the AUC and TR were very similar for the rotation estimator with $p = 0.5$ and $p = 1$. The AUC of the LASSO estimator with a small tuning parameter is similar to that of the $L^1$ rotation method. %We note that the AUC equals 1 for the $L^p$-based estimators and the LASSO estimator with small $\gamma$ under the smaller setting when $N=\{800, 1,600 \}$, and as well for the larger setting when $N = 1,600$. 
We noted that the model selection performance was poor for the LASSO estimator when $\gamma$ became large. This is due to the presence of many false negative selections (i.e., non-zero loading parameters selected as zeros), as a result of over-regularisation.

\begin{table}[h]
\centering
\caption{The AUC, TR, TPR, and TNR for the $L^p$-based rotation estimator and the regularised estimator, Study I.}
\label{tab:modacc}
\begin{tabular}{lcccccccc}
\hline
                                            & \multicolumn{4}{c|}{$15 \times 3$}                 & \multicolumn{4}{c}{$30 \times 5$} \\ \hline
                                            & AUC   & TR    & TPR   & \multicolumn{1}{c|}{TNR}   & AUC    & TR     & TPR    & TNR    \\ \hline
                                            & \multicolumn{8}{c}{$N=400$}                                                            \\ \hline
\multicolumn{1}{l|}{$L^{0.5}$ rotation}     & 0.996 & 0.979 & 0.979 & \multicolumn{1}{c|}{0.978} & 0.988  & 0.964  & 0.937  & 0.977  \\
\multicolumn{1}{l|}{$L^1$ rotation}         & 0.997 & 0.979 & 0.979 & \multicolumn{1}{c|}{0.979} & 0.988  & 0.964  & 0.937  & 0.977  \\
\multicolumn{1}{l|}{LASSO, $\gamma = 0.01$} & 0.997 & 0.981 & 0.979 & \multicolumn{1}{c|}{0.983} & 0.989  & 0.967  & 0.941  & 0.979  \\
\multicolumn{1}{l|}{LASSO, $\gamma = 0.05$} & 0.997 & 0.984 & 0.979 & \multicolumn{1}{c|}{0.988} & 0.992  & 0.971  & 0.944  & 0.985  \\
\multicolumn{1}{l|}{LASSO, $\gamma = 0.1$}  & 0.992 & 0.983 & 0.973 & \multicolumn{1}{c|}{0.991} & 0.987  & 0.970  & 0.937  & 0.986  \\
\multicolumn{1}{l|}{LASSO, $\gamma = 0.2$}  & 0.869 & 0.848 & 0.784 & \multicolumn{1}{c|}{0.905} & 0.903  & 0.917  & 0.800  & 0.975  \\
\multicolumn{1}{l|}{LASSO, $\gamma = 0.5$}  & 0.500 & 0.534 & 0.001 & \multicolumn{1}{c|}{1.000} & 0.520  & 0.683  & 0.075  & 0.987  \\ \hline
                                            & \multicolumn{8}{c}{$N=800$}                                                            \\ \hline
\multicolumn{1}{l|}{$L^{0.5}$ rotation}     & 1.000 & 0.993 & 0.992 & \multicolumn{1}{c|}{0.993} & 0.998  & 0.986  & 0.980  & 0.990  \\
\multicolumn{1}{l|}{$L^1$ rotation}         & 1.000 & 0.993 & 0.993 & \multicolumn{1}{c|}{0.992} & 0.998  & 0.987  & 0.981  & 0.990  \\
\multicolumn{1}{l|}{LASSO, $\gamma = 0.01$} & 1.000 & 0.992 & 0.992 & \multicolumn{1}{c|}{0.993} & 0.998  & 0.989  & 0.983  & 0.992  \\
\multicolumn{1}{l|}{LASSO, $\gamma = 0.05$} & 1.000 & 0.993 & 0.992 & \multicolumn{1}{c|}{0.993} & 0.999  & 0.990  & 0.985  & 0.993  \\
\multicolumn{1}{l|}{LASSO, $\gamma = 0.1$}  & 0.996 & 0.992 & 0.988 & \multicolumn{1}{c|}{0.995} & 0.995  & 0.989  & 0.979  & 0.994  \\
\multicolumn{1}{l|}{LASSO, $\gamma = 0.2$}  & 0.880 & 0.862 & 0.816 & \multicolumn{1}{c|}{0.902} & 0.919  & 0.932  & 0.824  & 0.987  \\
\multicolumn{1}{l|}{LASSO, $\gamma = 0.5$}  & 0.500 & 0.533 & 0.000 & \multicolumn{1}{c|}{1.000} & 0.506  & 0.672  & 0.024  & 0.996  \\ \hline
                                            & \multicolumn{8}{c}{$N=1,600$}                                                           \\ \hline
\multicolumn{1}{l|}{$L^{0.5}$ rotation}     & 1.000 & 0.997 & 0.999 & \multicolumn{1}{c|}{0.995} & 1.000  & 0.996  & 0.996  & 0.996  \\
\multicolumn{1}{l|}{$L^1$ rotation}         & 1.000 & 0.997 & 0.999 & \multicolumn{1}{c|}{0.995} & 1.000  & 0.996  & 0.996  & 0.996  \\
\multicolumn{1}{l|}{LASSO, $\gamma = 0.01$} & 1.000 & 0.997 & 1.000 & \multicolumn{1}{c|}{0.995} & 1.000  & 0.996  & 0.997  & 0.996  \\
\multicolumn{1}{l|}{LASSO, $\gamma = 0.05$} & 1.000 & 0.997 & 0.999 & \multicolumn{1}{c|}{0.996} & 1.000  & 0.997  & 0.998  & 0.996  \\
\multicolumn{1}{l|}{LASSO, $\gamma = 0.1$}  & 0.998 & 0.997 & 0.995 & \multicolumn{1}{c|}{0.999} & 0.998  & 0.996  & 0.993  & 0.997  \\
\multicolumn{1}{l|}{LASSO, $\gamma = 0.2$}  & 0.886 & 0.870 & 0.831 & \multicolumn{1}{c|}{0.904} & 0.929  & 0.941  & 0.838  & 0.993  \\
\multicolumn{1}{l|}{LASSO, $\gamma = 0.5$}  & 0.500 & 0.533 & 0.000 & \multicolumn{1}{c|}{1.000} & 0.501  & 0.668  & 0.003  & 0.999  \\ \hline
\end{tabular}
\end{table}

\begin{figure}
\begin{subfigure}{\textwidth} 
\includegraphics[width=\textwidth]{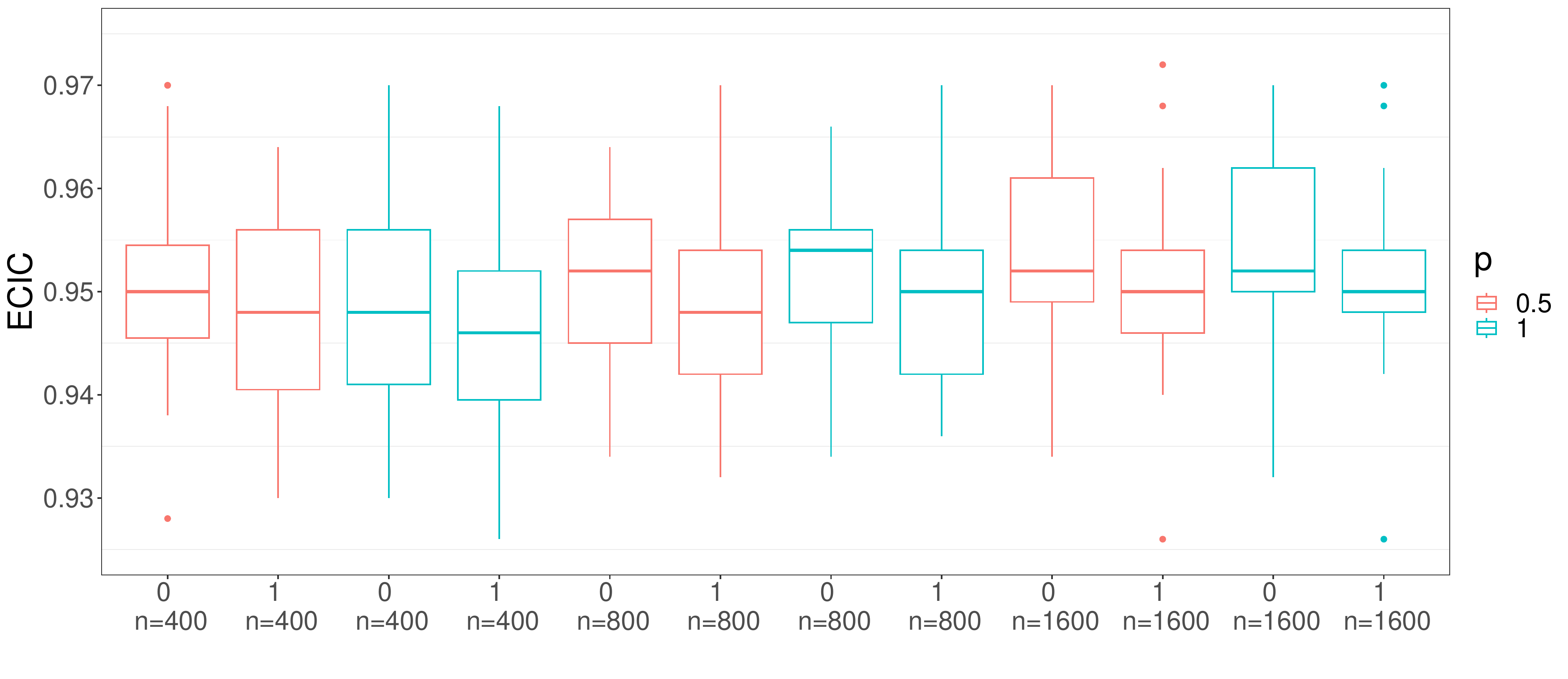}
\caption{Boxplots of $\mbox{ECIC}_{jk}$ for the $15 \times 3$ setting}
\end{subfigure}
\newline
\begin{subfigure}{\textwidth} 
\includegraphics[width=\textwidth]{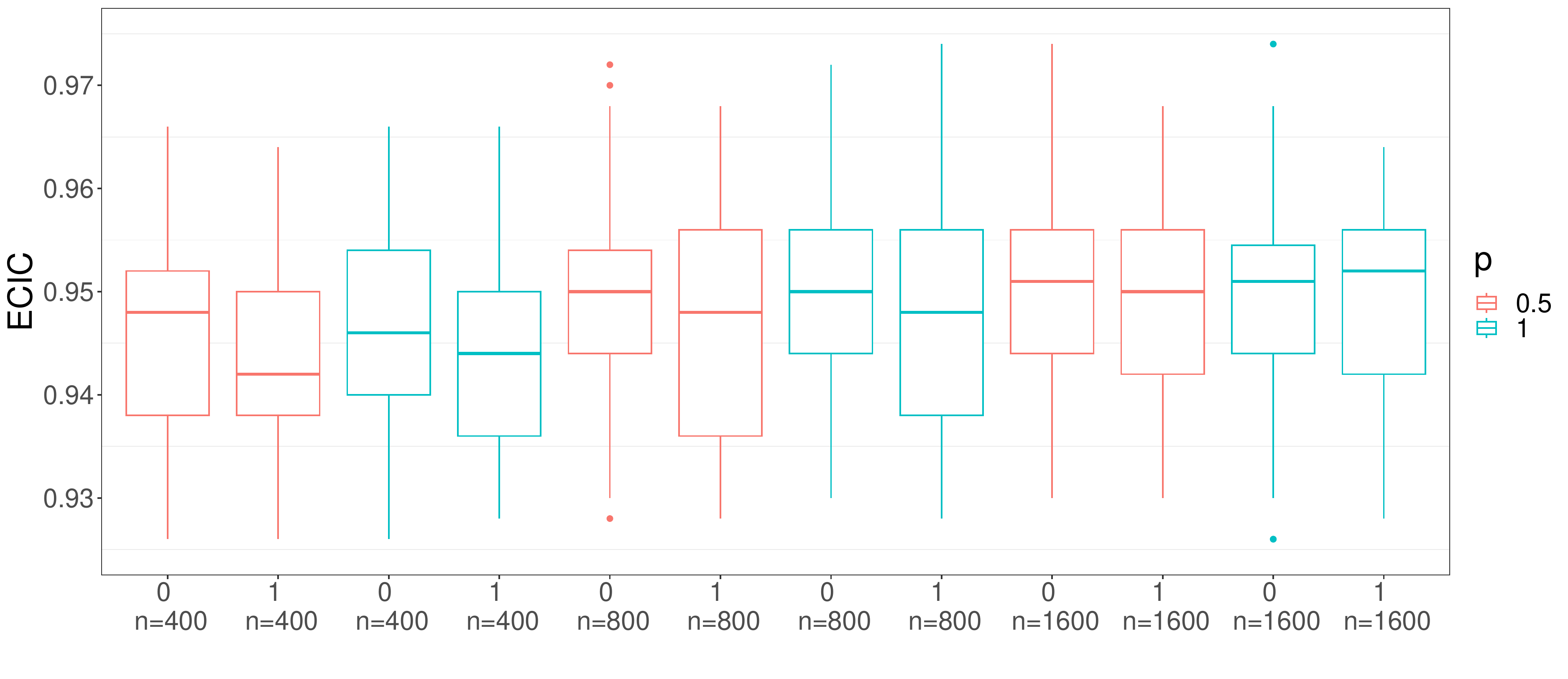}
\caption{Boxplots of $\mbox{ECIC}_{jk}$ for the $30 \times 5$ setting}
\end{subfigure}
       \caption{Boxplots of $\mbox{ECIC}_{jk}$. The label 0 means that $\lambda_{jk}^*=0$ and the label 1 means that $\lambda_{jk}^*\neq 0$.}
        \label{fig:ECIC}
\end{figure}

\paragraph{Results on confidence intervals.} In Figure \ref{fig:ECIC}, we show boxplots of the ECIC for the $L^p$ rotations, for $p = 0.5$ and $p = 1$ and  $N \in \{ 400, 800, 1,600 \}$. For both $p = 0.5$ and $p = 1$, the 
$\mbox{ECIC}_{jk}$s are close to the 95\% nominal level, supporting the consistency of the proposed procedure for constructing confidence intervals.

% is in the interval of 0.93-0.96 with smaller variation and higher median for larger sample sizes. When $N=1,600$, the median of $\mbox{ECIC}_{ij}$ in both settings achieve 95\%, for both zero entries and non-zero entries. This suggests that our post-selection inference method is valid for larger samples. \

\paragraph{Some remarks.} \yc{The computation  for the proposed $L^p$ rotation is fast. On a single core of a data science workstation,\footnote{CPU configuration: Intel Xeon 6246R 3.4GHz 2933MHz.} the mean time for solving the $L^1$ rotation criterion  is within  $0.29s$ for the $15\times 3 $ settings  and within $0.54s$ for the $30\times 5$ settings. Using the $L^{1}$ solution as the starting point, the mean time for solving the $L^{0.5}$ criterion
is within $0.13s$   for $15\times 3 $ settings and within $0.36s$ for the $30 \times 5 $ settings.} Under the current simulation settings, condition C3 is satisfied by both the $L^{0.5}$  and  $L^1$ criteria, in which cases the two criteria tend to perform similarly.  As we will show in Section~\ref{subsec:study2} below, the performance of the two criteria can be substantially different when C3 holds for one criterion but not the other. 
In addition, we see that the LASSO estimator with a small tuning parameter performed similarly to the $L^1$ rotation method. We expected this, since the 
$L^1$ rotation solution can be viewed as the limiting case of the LASSO estimator when the tuning parameter goes to zero. The LASSO estimator performed poorly for large tuning parameters, due to the bias brought by the regularisation. This bias-variance trade-off is visualised in \yc{Figure~\ref{fig:path}.  The two panels in Figure~\ref{fig:path} correspond to the $15\times 3$ and $30\times 5$ loading matrix settings, respectively.
For each panel, the $x$-axis shows the  tuning parameter $\gamma$, and the $y$-axis shows the MSE (for the loading matrix) of the corresponding LASSO estimator. The dots at $\gamma = 0$ correspond to the $L^1$ rotation solutions, as the $L^1$-rotation estimator is the limit of 
the LASSO estimator when $\gamma$ converges to zero (see Proposition~\ref{prop:regul}). As $\gamma$ increases, the estimation bias increases, and the variance decreases, which results in a U-shaped curve for the MSE -- a well-known phenomenon in statistical learning theory \citep[see Chapter 2,][]{hastie2009elements}. However, the U-shaped curves in Figure~\ref{fig:path} are very asymmetric -- the MSE only decreases slightly before increasing. This means that the estimators with small $\gamma$ values including the rotation solution have similar estimation accuracy to the optimal choice of the tuning parameter (i.e., the value of $\gamma$ at which the MSE curve achieves the minimum value). In that case, it may not be worth searching for the optimal tuning parameter, as constructing a LASSO solution path is typically computationally intensive. Instead, using the rotation method or a LASSO estimator with a sufficiently small tuning parameter is computationally more affordable and yields a sufficiently accurate solution.}

%Finally, it is not clear to us why the geomin rotation performs similarly as the $L^{0.5}$ and  $L^1$ criteria, which is worth future investigation. 

\begin{comment}
\begin{figure}[h]
\includegraphics[width=\textwidth]{510_30_5_400.png}
\end{figure}

\begin{figure}[h]
\includegraphics[width=\textwidth]{510_30_5_800.png}
\end{figure}

\begin{figure}[h]
\includegraphics[width=\textwidth]{510_30_5_1,600.png}
\end{figure}
\end{comment}

\subsection{Study II}\label{subsec:study2}

In this study we compare the $L^{0.5}$ and $L^1$ rotations, 
under a setting where condition C3 holds for the $L^{0.5}$ rotation but not the $L^1$ rotation. We chose the setting to somewhat exaggerate the differences, in order to show  the consequence of misspecifying $p$.

\paragraph{Setting and evaluation criteria.} %~\yc{Describe the setting and evaluation criteria. The true model parameters should be given in the supplement. Here, only describe the setting of the parameters. }  
The true loading matrix is of dimension $J = 18$ and $K = 3$.
Each 
%to be  less sparse than the true loading matrix in Study I.
item is set to load on two factors, so that no item has a perfect simple structure. 
Given the loading structure, the model is identifiable as a confirmatory factor analysis model. We present the true model parameters in the supplementary material. By grid search, we checked that C3 holds for the $L^{0.5}$ criterion but not the $L^{1}$ criterion. We chose the sample size to be $N = 3,000$. Similar to Study I, we compare the two rotation criteria using the MSE, AUC, TR, TPR, and TNR by running $B = 500 $ independent replications.  

\paragraph{Results.} %~\yc{Present and discuss the results.}
We present the results in Table \ref{tab:counterex}. 
%, the MSE, AUC, TR, TPR and TNR are presented for the $L^{0.5}$- and $L^1$-rotation methods. 
The $L^{0.5}$ criterion performed better in terms of both point estimation and model selection, as its MSE was lower and the AUC, TR, TPR, and TNR were higher. 
%The two criteria showed a clear difference in TNR, 
%for which the $L^{0.5}$-criterion is as high as 0.967 while the 
In particular, we noted that the $L^{0.5}$ rotation achieved a much higher TNR than the $L^1$ rotation, meaning that the 
$L^1$ rotation tended to make many false positive selections (i.e., zero loading parameters selected as non-zeros), as a consequence of violating condition C3.

%which is an improvement of 0.255. This suggests $L^1$ produce much less zeros compared to the true matrix. A true zero-loading would thus be discovered much more often using the $L^{0.5}$ loss, for this considered scenario.  

% \usepackage{graphicx}

\begin{table}[t]
\centering
\caption{The MSE, AUC, TR, TPR, and TNR for the $L^{p}$-based rotation estimator, Study II.}
\label{tab:counterex}
\resizebox{0.5\textwidth}{!}{%
\begin{tabular}{lccccc}
\hline
                             & MSE   & AUC   & TR & TPR &TNR \\ \hline
\multicolumn{1}{l|}{$L^{0.5}$ rotation} & 0.003 & 0.984 & 0.954  & 0.943   & 0.974   \\
\multicolumn{1}{l|}{$L^1$ rotation}   & 0.025 & 0.953 & 0.865  & 0.936   & 0.725   \\
 \hline
\end{tabular}%
}
\end{table}

\section{An Application to the Big Five Personality Test} \label{sec:real}

We illustrate the proposed method through an application to the Big Five personality test. We consider the Big Five Factor Markers  from the International Personality Item Pool \citep[][]{goldberg1992development}, which contains 50 items designed to measure five personality factors, namely Extraversion (E), Emotional Stability (ES), Agreeableness (A), Conscientiousness (C), and Intellect/Imagination (I). Each item is a statement describing a personality pattern like "\textit{I am the life of the party}" and "\textit{I get stressed out easily}", designed to primarily measure one personality factor.  
We can divide the 50 items into five equal-sized groups, with each group mainly measuring one personality factor. 
Responses to the items are on  a five-level Likert scale, which we treat as continuous variables in the current analysis. 

Although the Big Five personality test was designed to have a perfect simple structure, cross-loadings are often found in empirical data \citep[e.g.,][]{gow2005goldberg}. To better understand the loading structure of this widely used personality test, we applied the proposed $L^{0.5}$  and $L^1$ rotations to a dataset\footnote{The dataset is downloaded from: \url{http://personality-testing.info/_rawdata/}.} on this test. To avoid possible complexities brought by measurement non-invariance, we selected the subset of male respondents from the United Kingdom, which has a sample size  $N = 609$. In the analysis, the number of factors is set to be $K=5$.

After applying the proposed rotations, we further adjusted the estimates by column permutation and sign flip transformations, so that the resulting factors correspond to the E, ES, A, C, and I factors, respectively. 
We give our results in Tables~\ref{tab:empcovmat} through \ref{tab:empCI0.5_3}. In Table \ref{tab:empcovmat} we show the estimated covariance matrices from the two rotations. The estimated correlation matrices from the two criteria are similar to each other. In particular, all the signs of the correlations are consistent, except for the correlation between A and I, in which case both correlations are close to zero.  In addition,
for each pair of factors, the correlations obtained by the two criteria are close. The sign pattern of 
the correlations between the Big Five factors is largely consistent with those found in the literature \citep[e.g.,][]{booth2014exploratory,gow2005goldberg}. 

In Tables \ref{tab:empCI0.5_1} through \ref{tab:empCI0.5_3} we show the estimated loading parameters and the corresponding 95\% confidence intervals obtained from the $L^{0.5}$ rotation. 
We indicate by asterisks the loadings that are significantly different from zero according to the 95\% confidence intervals. The results of the $L^1$ rotation are similar and thus we give them in the supplementary material. In Tables \ref{tab:empCI0.5_1}-\ref{tab:empCI0.5_3}, the items are labelled based on the personality factor that they are designed to measure, and their scoring keys.\footnote{Positively scored items are indicated by ``$(+)$" and negatively scored items are indicated by ``$(-)$".} The estimated loading matrix is largely consistent with the International Personality Item Pool \yc{(IPIP) scoring key}, where all the items have relatively strong loadings on the factors that they are designed to measure, and the signs of the loadings are consistent with the scoring keys.  The confidence intervals shed additional light on the uncertainty of each loading. Specifically, we notice that many loadings are statistically insignificantly different from zero, suggesting that the true loading structure is sparse.  There are also items with fairly strong cross-loadings. 

\begin{table}[h]
\centering
\caption{Estimated correlation matrices based on $L^{0.5}$ and $L^1$ rotations, Big Five personality test}
\label{tab:empcovmat}
\begin{tabular}{lccccc|ccccc}
\hline
\multicolumn{1}{c}{}    & \multicolumn{5}{c|}{$p=0.5$}         & \multicolumn{5}{c}{$p=1$}           \\ \hline
\multicolumn{1}{c}{}    & E     & ES     & A      & C      & I & E     & ES     & A     & C      & I \\ \hline
\multicolumn{1}{l|}{E}  & 1     &        &        &        &   & 1     &        &       &        &   \\
\multicolumn{1}{l|}{ES} & 0.154 & 1      &        &        &   & 0.184 & 1      &       &        &   \\
\multicolumn{1}{l|}{A}  & 0.193 & -0.017 & 1      &        &   & 0.197 & -0.001 & 1     &        &   \\
\multicolumn{1}{l|}{C}  & 0.016 & 0.010  & 0.023  & 1      &   & 0.022 & 0.148  & 0.038 & 1      &   \\
\multicolumn{1}{l|}{I}  & 0.050 & 0.018   & -0.005 & -0.046 & 1 & 0.161 & 0.040  & 0.023  & -0.019 & 1 \\ \hline
\end{tabular}
\end{table}

% \usepackage{graphicx}

% latex table generated in R 4.2.2 by xtable 1.8-4 package
% Wed Nov 23 10:45:57 2022
\begin{table}
\centering
\caption{ Part I: Point estimates and confidence intervals constructed by $L^{0.5}$, Big Five personality test. The loadings that are significantly different from zero according to the 95\% confidence intervals are indicated by asterisks.} 
\begin{tabular}{l|ccccc}
  \hline
  & E & ES & A & C & I \\ 
  \hline
E1(+) &  0.887* & -0.069  & -0.068* &  0.004  &  0.066  \\ 
    & ( 0.795, 0.984) & (-0.158, 0.005) & (-0.182,-0.016) & (-0.113, 0.059) & (-0.030, 0.140) \\ 
  E2(-) & -0.851* &  0.131* &  0.003  &  0.047  &  0.001  \\ 
    & (-0.969,-0.765) & ( 0.049, 0.228) & (-0.057, 0.126) & (-0.021, 0.168) & (-0.071, 0.118) \\ 
  E3(+) &  0.780* &  0.276* &  0.204* &  0.142* & -0.107* \\ 
    & ( 0.703, 0.879) & ( 0.190, 0.344) & ( 0.121, 0.277) & ( 0.060, 0.220) & (-0.200,-0.042) \\ 
  E4(-) & -0.914* & -0.058  & -0.022  &  0.002  &  0.105* \\ 
    & (-1.022,-0.844) & (-0.139, 0.012) & (-0.075, 0.077) & (-0.066, 0.094) & ( 0.050, 0.205) \\ 
  E5(+) &  0.898* & -0.024  &  0.155* &  0.100  &  0.064  \\ 
    & ( 0.814, 0.991) & (-0.116, 0.034) & ( 0.060, 0.212) & (-0.001, 0.155) & (-0.016, 0.140) \\ 
  E6(-) & -0.754* & -0.001  & -0.088  & -0.061  & -0.123* \\ 
    & (-0.854,-0.662) & (-0.066, 0.106) & (-0.163, 0.010) & (-0.152, 0.027) & (-0.200,-0.023) \\ 
  E7(+) &  1.119* & -0.078* &  0.083* &  0.092* & -0.042  \\ 
    & ( 1.025, 1.228) & (-0.187,-0.019) & ( 0.002, 0.174) & ( 0.005, 0.184) & (-0.175, 0.002) \\ 
  E8(-) & -0.724* & -0.086  &  0.028  &  0.115* & -0.051  \\ 
    & (-0.829,-0.634) & (-0.173, 0.002) & (-0.036, 0.142) & ( 0.026, 0.208) & (-0.129, 0.056) \\ 
  E9(+) &  0.862* &  0.051  &  0.000  & -0.010  &  0.226* \\ 
    & ( 0.751, 0.958) & (-0.048, 0.136) & (-0.110, 0.075) & (-0.127, 0.067) & ( 0.110, 0.301) \\ 
  E10(-) & -0.828* & -0.117* & -0.049  & -0.126* &  0.020  \\ 
    & (-0.935,-0.745) & (-0.189,-0.021) & (-0.124, 0.046) & (-0.212,-0.036) & (-0.043, 0.132) \\ 
  ES1(-) & -0.132* & -0.971* &  0.006  &  0.001  & -0.101  \\ 
    & (-0.215,-0.028) & (-1.065,-0.869) & (-0.117, 0.082) & (-0.133, 0.054) & (-0.175, 0.003) \\ 
  ES2(+) &  0.147* &  0.671* &  0.001  & -0.029  &  0.082  \\ 
    & ( 0.039, 0.220) & ( 0.587, 0.768) & (-0.066, 0.112) & (-0.113, 0.064) & (-0.008, 0.170) \\ 
  ES3(-) & -0.186* & -0.780* &  0.231* &  0.063  &  0.046  \\ 
    & (-0.277,-0.095) & (-0.880,-0.696) & ( 0.128, 0.306) & (-0.041, 0.138) & (-0.008, 0.170) \\ 
  ES4(+) &  0.225* &  0.565* &  0.002  &  0.110* &  0.006  \\ 
    & ( 0.116, 0.314) & ( 0.468, 0.664) & (-0.071, 0.122) & ( 0.024, 0.224) & (-0.105, 0.090) \\ 
  ES5(-) &  0.013  & -0.473* & -0.042  & -0.152* & -0.226* \\ 
    & (-0.075, 0.137) & (-0.566,-0.356) & (-0.163, 0.046) & (-0.272,-0.059) & (-0.337,-0.125) \\ 
  ES6(-) & -0.130* & -0.806* &  0.257* & -0.088* & -0.130* \\ 
    & (-0.205,-0.023) & (-0.903,-0.716) & ( 0.147, 0.328) & (-0.223,-0.039) & (-0.209,-0.033) \\ 
  ES7(-) &  0.022  & -0.962* & -0.112* & -0.124* &  0.004  \\ 
    & (-0.051, 0.119) & (-1.051,-0.867) & (-0.224,-0.050) & (-0.244,-0.064) & (-0.073, 0.089) \\ 
  ES8(-) &  0.000  & -1.131* & -0.135* & -0.169* &  0.000  \\ 
    & (-0.085, 0.100) & (-1.227,-1.029) & (-0.258,-0.075) & (-0.294,-0.103) & (-0.078, 0.095) \\ 
  ES9(-) & -0.033  & -0.862* & -0.293* &  0.097  & -0.016  \\ 
    & (-0.134, 0.048) & (-0.949,-0.764) & (-0.394,-0.211) & (-0.002, 0.183) & (-0.095, 0.082) \\ 
  ES10(-) & -0.344* & -0.837* &  0.069  & -0.172* &  0.104* \\ 
    & (-0.439,-0.256) & (-0.930,-0.742) & (-0.026, 0.157) & (-0.284,-0.101) & ( 0.032, 0.206) \\ 
   \hline
\end{tabular}
\label{tab:empCI0.5_1}
\end{table}
% latex table generated in R 4.2.2 by xtable 1.8-4 package
% Wed Nov 23 10:45:57 2022
\begin{table} 
\centering
\caption{ Part II: Point estimates and confidence intervals constructed by $L^{0.5}$, Big Five personality test.} 
\begin{tabular}{l|ccccc}
  \hline
  & E & ES & A & C & I \\ 
  \hline
A1(-) &  0.003  & -0.127* & -0.778* &  0.010  &  0.045  \\ 
    & (-0.114, 0.087) & (-0.201,-0.011) & (-0.875,-0.669) & (-0.095, 0.103) & (-0.060, 0.136) \\ 
  A2(+) &  0.439* & -0.007  &  0.557* & -0.038  &  0.035  \\ 
    & ( 0.361, 0.526) & (-0.097, 0.054) & ( 0.464, 0.626) & (-0.132, 0.024) & (-0.042, 0.113) \\ 
  A3(-) &  0.193* & -0.575* & -0.566* & -0.130* &  0.134* \\ 
    & ( 0.080, 0.286) & (-0.663,-0.456) & (-0.691,-0.479) & (-0.265,-0.054) & ( 0.026, 0.230) \\ 
  A4(+) &  0.013  &  0.001  &  0.979* & -0.002  & -0.002  \\ 
    & (-0.035, 0.144) & (-0.102, 0.038) & ( 0.895, 1.050) & (-0.081, 0.047) & (-0.051, 0.083) \\ 
  A5(-) & -0.155* & -0.039  & -0.815* & -0.012  &  0.090* \\ 
    & (-0.250,-0.091) & (-0.102, 0.049) & (-0.894,-0.724) & (-0.085, 0.069) & ( 0.017, 0.170) \\ 
  A6(+) & -0.059  & -0.182* &  0.717* &  0.002  &  0.014  \\ 
    & (-0.159, 0.020) & (-0.272,-0.105) & ( 0.629, 0.811) & (-0.101, 0.070) & (-0.063, 0.110) \\ 
  A7(-) & -0.367* & -0.089* & -0.733* &  0.043  &  0.036  \\ 
    & (-0.456,-0.300) & (-0.159,-0.015) & (-0.800,-0.639) & (-0.023, 0.125) & (-0.032, 0.115) \\ 
  A8(+) &  0.111* & -0.038  &  0.692* &  0.085* &  0.025  \\ 
    & ( 0.039, 0.185) & (-0.128, 0.010) & ( 0.617, 0.771) & ( 0.010, 0.152) & (-0.035, 0.107) \\ 
  A9(+) &  0.123* & -0.110* &  0.751* &  0.066  &  0.110* \\ 
    & ( 0.040, 0.199) & (-0.204,-0.054) & ( 0.668, 0.836) & (-0.018, 0.137) & ( 0.041, 0.195) \\ 
  A10(+) &  0.439* &  0.071  &  0.321* &  0.133* &  0.045  \\ 
    & ( 0.354, 0.517) & (-0.010, 0.143) & ( 0.245, 0.402) & ( 0.043, 0.201) & (-0.037, 0.121) \\ 
  C1(+) &  0.105  &  0.111  & -0.037  &  0.695* &  0.129* \\ 
    & (-0.001, 0.179) & (-0.005, 0.177) & (-0.099, 0.088) & ( 0.597, 0.785) & ( 0.055, 0.238) \\ 
  C2(-) &  0.080  & -0.201* &  0.107  & -0.670* &  0.142* \\ 
    & (-0.014, 0.194) & (-0.284,-0.073) & (-0.028, 0.177) & (-0.800,-0.585) & ( 0.013, 0.217) \\ 
  C3(+) &  0.023  &  0.007  &  0.114* &  0.407* &  0.280* \\ 
    & (-0.080, 0.082) & (-0.094, 0.065) & ( 0.050, 0.210) & ( 0.315, 0.482) & ( 0.213, 0.378) \\ 
  C4(-) & -0.123* & -0.613* &  0.048  & -0.544* & -0.039  \\ 
    & (-0.202,-0.036) & (-0.671,-0.495) & (-0.057, 0.114) & (-0.656,-0.483) & (-0.149, 0.018) \\ 
  C5(+) &  0.074  &  0.057  &  0.000  &  0.782* & -0.052  \\ 
    & (-0.005, 0.188) & (-0.051, 0.158) & (-0.041, 0.163) & ( 0.687, 0.882) & (-0.133, 0.061) \\ 
  C6(-) &  0.021  & -0.195* &  0.045  & -0.718* &  0.087  \\ 
    & (-0.085, 0.130) & (-0.276,-0.058) & (-0.090, 0.128) & (-0.848,-0.625) & (-0.028, 0.188) \\ 
  C7(+) & -0.129* & -0.128* &  0.110* &  0.520* &  0.042  \\ 
    & (-0.225,-0.047) & (-0.236,-0.059) & ( 0.041, 0.220) & ( 0.427, 0.608) & (-0.015, 0.166) \\ 
  C8(-) & -0.000  & -0.284* & -0.242* & -0.549* & -0.000  \\ 
    & (-0.086, 0.097) & (-0.349,-0.166) & (-0.361,-0.179) & (-0.654,-0.466) & (-0.119, 0.063) \\ 
  C9(+) &  0.031  & -0.003  &  0.123* &  0.722* & -0.076  \\ 
    & (-0.061, 0.129) & (-0.140, 0.060) & ( 0.059, 0.248) & ( 0.623, 0.816) & (-0.157, 0.034) \\ 
  C10(+) & -0.001  & -0.006  &  0.127* &  0.528* &  0.235* \\ 
    & (-0.110, 0.057) & (-0.120, 0.046) & ( 0.070, 0.236) & ( 0.433, 0.605) & ( 0.170, 0.338) \\ 
   \hline
\end{tabular}

\label{tab:empCI0.5_2}
\end{table}
% latex table generated in R 4.2.2 by xtable 1.8-4 package
% Wed Nov 23 10:45:57 2022
\begin{table}
\centering
\caption{ Part III: Point estimates and confidence intervals constructed by $L^{0.5}$, Big Five personality test.} 
\begin{tabular}{l|ccccc}
  \hline
  & E & ES & A & C & I \\ 
  \hline
I1(+) &  0.085  &  0.001  & -0.046  &  0.009  &  0.621* \\ 
    & (-0.020, 0.145) & (-0.102, 0.067) & (-0.148, 0.014) & (-0.088, 0.078) & ( 0.537, 0.713) \\ 
  I2(-) & -0.000  & -0.222* & -0.087* & -0.020  & -0.581* \\ 
    & (-0.052, 0.120) & (-0.289,-0.119) & (-0.183,-0.014) & (-0.103, 0.071) & (-0.675,-0.498) \\ 
  I3(+) &  0.076  & -0.152* &  0.024  & -0.000  &  0.587* \\ 
    & (-0.028, 0.137) & (-0.244,-0.087) & (-0.061, 0.095) & (-0.103, 0.063) & ( 0.503, 0.670) \\ 
  I4(-) &  0.023  & -0.204* & -0.154* &  0.008  & -0.572* \\ 
    & (-0.027, 0.145) & (-0.269,-0.101) & (-0.228,-0.062) & (-0.074, 0.098) & (-0.663,-0.487) \\ 
  I5(+) &  0.240* &  0.068* & -0.058  &  0.189* &  0.575* \\ 
    & ( 0.131, 0.272) & ( 0.003, 0.139) & (-0.130, 0.002) & ( 0.095, 0.240) & ( 0.501, 0.648) \\ 
  I6(-) & -0.217* & -0.001  & -0.047  &  0.020  & -0.505* \\ 
    & (-0.275,-0.104) & (-0.065, 0.102) & (-0.120, 0.046) & (-0.053, 0.119) & (-0.597,-0.421) \\ 
  I7(+) &  0.076  &  0.168* & -0.035  &  0.117* &  0.449* \\ 
    & (-0.018, 0.123) & ( 0.089, 0.226) & (-0.100, 0.036) & ( 0.032, 0.176) & ( 0.376, 0.520) \\ 
  I8(+) & -0.014  & -0.163* & -0.108* & -0.003  &  0.656* \\ 
    & (-0.163, 0.023) & (-0.262,-0.082) & (-0.198,-0.020) & (-0.130, 0.058) & ( 0.572, 0.769) \\ 
  I9(+) & -0.056  & -0.213* &  0.239* &  0.097* &  0.260* \\ 
    & (-0.153, 0.008) & (-0.305,-0.148) & ( 0.157, 0.317) & ( 0.006, 0.167) & ( 0.188, 0.350) \\ 
  I10(+) &  0.246* & -0.004  & -0.000  &  0.107* &  0.680* \\ 
    & ( 0.130, 0.276) & (-0.108, 0.039) & (-0.072, 0.067) & ( 0.009, 0.159) & ( 0.606, 0.761) \\ 
   \hline
\end{tabular}

\label{tab:empCI0.5_3}
\end{table} 

\section{Concluding Remarks}\label{sec:conc}
 
In this paper we propose a new family of oblique 
rotations based on component-wise $L^p$ loss functions $(0 < p\leq 1)$ and establish the relationship between the proposed rotation estimator and the $L^p$ regularised estimator for EFA. We develop point estimation, model selection, and post-selection inference procedures and establish their asymptotic theories. We also develop an iteratively reweighted gradient projection algorithm for the computation\footnote{The R code for the proposed method is available from \url{https://github.com/yunxiaochen/Lp_rot1129}}. We demonstrate the power of the proposed method via simulation studies and an application to Big Five personality assessment. 

We note that the proposed procedures do not rely on the normality assumption of the EFA model, though we make such an assumption in the problem setup for ease of exposition. Specifically, in the rotation, we only need to obtain a consistent initial estimator for EFA in the sense of  condition C1, which we can obtain with any reasonable loss function for factor analysis. In the model selection, only the BIC uses the likelihood function based on the normal model. Note that the likelihood function is a valid loss function under the linear factor model, even if the normality assumption does not hold \citep[Chapter 7,][]{bollen1989structural}. Therefore, the BIC still yields consistent model selection under the misspecification of the normality assumption \citep{machado1993robust}. Finally, 
the confidence intervals are based on the asymptotic distributions of CFA models. If we use a robust method (i.e., a sandwich estimator) for computing the asymptotic variance, then the resulting confidence intervals are valid when the normality assumption does not hold.

\yc{As each value of $p \in (0, 1]$ leads to a sensible rotation criterion, which $L^p$ criterion should we use? We do not recommend trying too many values of $p$.} From the previous discussion, we see that there is a statistical and computational trade-off underlying the choice of $p$. Theoretically, a smaller value of $p$ is more likely to recover a sparse loading matrix, but the associated optimisation problem is computationally more challenging. \yc{The $L^1$ criterion is the easiest to compute. Although we gave an example earlier in which the $L^1$ criterion fails to recover the sparest loading structure, the $L^1$ criterion can accurately recover the true loading structure under most simulation settings. For several real-world datasets we have encountered, different $p$ values also give very similar results. We thus believe that the $L^1$ criterion is robust and recommend users to always start with the $L^1$ criterion. 
To check the result of the $L^1$ criterion, users may try some smaller $p$ values (e.g., $p=0.5$) and compare their results with the $L^1$ result in terms of model fitting and substantive interpretations. If they give similar results, then the best fitting solution should be reported. If the result from a smaller $p$ value substantially differs from the $L^1$ result, then the value of $p$ should be further decreased until the result stabilises. Computationally, when solving the optimisation with a smaller value of $p$, we recommend using the solution from the previous larger value of $p$ as the starting point, so that the algorithm is less likely to get stuck at a local optimum.

}

%Our recommendation is to try multiple values of $p$, starting from $p=1$ and gradually decreasing the value of $p$. Computationally, when solving the optimisation with a smaller value of $p$, we recommend using the solution from the previous larger value of $p$ as the starting point, so that the algorithm is less likely to get stuck at a local optimum. Then, one may compare these solutions based on their model selection results using the BIC or other criteria for model comparison. 

%\clearpage

%\yx{In this paper, we are considering $p$ to be fixed. Statistically speaking, the so-called $L^0$ norm, i.e., the number of non-zero elements in the estimated loading matrix is optimal as it corresponds to best subset selection.} 

Our complexity analysis and simulation results suggest that obtaining a solution path for the $L^1$-regularised estimator has little added value over the $L^1$ rotation  when the sample size is reasonably large. That is, obtaining the solution path of the regularised estimator is computationally more intensive, while the best tuning parameter is often very close to zero and thus the corresponding solution is very similar to the rotation solution. Therefore, when the sample size is reasonably large, we do not recommend running a solution path for the $L^1$ regularised estimator to learn the loading structure in EFA. Instead, users can obtain a point estimate by either applying the $L^1$ rotation or running the $L^1$ regularised estimator with a single small tuning parameter. Model selection can be done by applying hard-thresholding to this point estimate. Furthermore, although an $L^p$ regularised estimator is mathematically well-defined with $p<1$, algorithms remain to be developed for its computation. On the other hand, $L^p$ rotation can be computed by the proposed IRGP algorithm for all $p \in (0,1]$. \yc{However, when the sample size is small and the number of items is large,  the regularised estimators may outperform their rotation counterparts. In that case, an optimally tuned regularised estimator may be substantially more accurate than those with very small tuning parameters or the rotation-based estimator, and thus, better learn the sparse loading structure.}

%\clearpage

%\yx{would like to point out that we do not discard $L^p$ regularisation. Our theoretical results, which establish the connection between the regularised estimator and the family of rotation criteria based on the $L^p$ loss function, however show that the computationally demanding search for an optimal weight for the regularised estimator can be bypassed by instead consider a rotation based on the $L^p$ loss. Lastly, allowing for $0 < p \leq 1$ permits more complex sparsity structures. } 

The current work has several limitations that require future investigation.  First, the way the confidence intervals are constructed may be improved. That is, accurate model selection (condition C5)
and identifiability conditions on the true model (condition C6) are required for the confidence intervals to have good coverage rate, while the uncertainty in the model selection step is not taken into account in the proposed procedure. Consequently, although the proposed confidence intervals are shown to be asymptotically valid, they may not perform well when the sample size is small. 
This issue may be addressed by future researchers developing bootstrap procedures for constructing confidence intervals,
as such procedures may still be valid even when the objective function is nonsmooth \citep{sen2010inconsistency}. 

The current theoretical results only consider a low-dimensional setting where the numbers of manifest variables and factors are fixed and the sample size goes to infinity. As factor analysis is commonly used by those analysing high-dimensional multivariate data, it is of interest to 
generalise the current results to a high-dimensional regime where the numbers of manifest variables, factors, and observations all grow to infinity \citep{chen2019joint,chen2020structured,zhang2020note,chendetermining}. In particular, it will be of interest to see how the rotation methods work with the joint maximum likelihood estimator for high-dimensional factor models \citep{chen2019joint,chen2020structured}.

Finally, as is an issue with any simulation study, we can only examine a small number of simulation settings, and thus, may not be able to provide a complete picture of the proposed methods. Future researchers need to investigate more simulation settings by varying the numbers of manifest variables, factors, and observations, the sign  pattern of the true loading matrix, and the generation mechanism of the true model parameters.

\bigskip\bigskip\bigskip
 
%\yx{Please check the references. Check the upper and lower cases. Volume, page, etc. Note that Psychometrika does not need number (i.e., the number in the brackets after volume).}
%\bibliography{bibliography}
\printbibliography
\end{document}